\documentclass[a4paper,fleqn,usenatbib]{mnras}
\usepackage{graphicx}

\newcommand{\vsini}{\mbox{$v \sin i$}}

\newcommand{\sqiglt}{\hbox{\rlap{\lower.55ex \hbox {$\sim$}}
	\kern-.3em \raise.4ex \hbox{$<$}\,}}
\newcommand{\sqiggt}{\hbox{\rlap{\lower.55ex \hbox {$\sim$}}
	\kern-.3em \raise.4ex \hbox{$>$}\,}}

\title[WASP-South transiting exoplanets]{New transiting hot Jupiters discovered by WASP-South, Euler/CORALIE and TRAPPIST-South}

\author[Hellier et al.]{Coel Hellier$^{1}$,
D.R. Anderson$^{1}$, 
F. Bouchy$^{2}$,
A. Burdanov$^{3}$,
A. Collier Cameron$^{4}$,  \newauthor  
L. Delrez$^{3,5}$,
M. Gillon$^{3}$,
E. Jehin$^{3}$, 
M. Lendl$^{6}$, 
L.D. Nielsen$^{2}$, 
P.F.L. Maxted$^{1}$, \newauthor    
F. Pepe$^{2}$, 
D. Pollacco$^{7}$,    
D. Queloz$^{5}$, 
D. S\'egransan$^{2}$, 
B. Smalley$^{1}$,  \newauthor 
A.H.M.J. Triaud$^{8}$,  
S. Udry$^{2}$, and 
R.G. West$^{7}$\\    
$^{1}$Astrophysics Group, Keele University, Staffordshire, ST5 5BG, UK\\
$^{2}$Observatoire astronomique de l'Universit\'e de Gen\`eve
51 ch. des Maillettes, 1290 Sauverny, Switzerland\\
$^{3}$Institut d'Astrophysique et de G\'eophysique, Universit\'e de
Li\`ege, All\'ee du 6 Ao\^ut, 17, Bat. B5C, Li\`ege 1, Belgium\\
$^{4}$SUPA, School of Physics and Astronomy, University of St.\ Andrews, North Haugh,  Fife, KY16 9SS, UK\\
$^{5}$Cavendish Laboratory, J J Thomson Avenue, Cambridge, CB3 0HE, UK\\
$^{6}$Space Research Institute, Austrian Academy of Sciences, Schmiedlstr. 6, 8042, Graz, Austria\\
$^{7}$Department of Physics, University of Warwick, Gibbet Hill Road, Coventry CV4 7AL, UK\\
$^{8}$School of Physics \&\ Astronomy, University of Birmingham, Edgbaston, Birmingham, B15 2TT, UK}

\begin{document}

\date{date}
\pagerange{range}

\maketitle

\begin{abstract}We report the discovery of eight hot-Jupiter exoplanets from the WASP-South transit survey.  \\
WASP-144b has a mass of  0.44 M$_{\rm Jup}$, a radius of 0.85 R$_{\rm Jup}$, and
 is  in a 2.27-d orbit around a $V$ = 12.9, K2 star which shows a 21-d rotational modulation. \\
WASP-145Ab is a 0.89 M$_{\rm Jup}$ planet in a 1.77-d orbit with a grazing transit.  The host is a $V$ = 11.5, K2 star with a companion 5 arcsecs away and 1.4 mags fainter.  \\
WASP-158b is a relatively massive planet at 2.8 M$_{\rm Jup}$ with a radius of 1.1 R$_{\rm Jup}$ and a 3.66-d orbit.  It transits a  $V$ = 12.1, F6 star. \\
WASP-159b is a bloated hot Jupiter (1.4 R$_{\rm Jup}$ and 0.55 M$_{\rm Jup}$) in a 3.8-d orbit around a $V$ = 12.9, F9 star.\\
WASP-162b is a massive planet in a relatively long and highly eccentric orbit 
(5.2 M$_{\rm Jup}$, $P$ = 9.6 d, $e$ = 0.43).   It transits a $V$ = 12.2, K0 star. \\
WASP-168b is a bloated hot Jupiter (0.42 M$_{\rm Jup}$, 1.5 R$_{\rm Jup}$) in a 4.15-d orbit with a grazing transit.  The host is a $V$ = 12.1, F9 star. \\
WASP-172b is a bloated hot Jupiter (0.5 M$_{\rm Jup}$; 1.6 R$_{\rm Jup}$) in a 5.48-d orbit around a $V$ = 11.0, F1 star.\\
WASP-173Ab is a massive planet (3.7 M$_{\rm Jup}$) with a 1.2 R$_{\rm Jup}$ radius in a circular orbit with a period of 1.39 d.  The host is a $V$ = 11.3, G3 star, being the brighter component of the double-star system WDS23366$-$3437, with a companion 6 arcsecs away and 0.8 mags fainter. One of the two stars shows a rotational modulation of 7.9 d.  
\end{abstract}


\section{Introduction}
Hot-Jupiter exoplanets are relatively rare, being found in only $\sim$\,1\%\ of Solar-like stars. However, since they are the easiest planets to detect they are by the far the commonest type of planet found by ground-based transit surveys such as WASP \citep{2006PASP..118.1407P}. Such planets produce relatively deep transits ($\sim$ 1\%) that recur often owing to short orbital periods (1--10 d).  A massive planet in a close-in orbit will also produce a radial-velocity signal large enough for verification with relatively small telescopes such as the Swiss Euler 1.2-m and its CORALIE spectrograph.    The ease of studying hot Jupiters also makes them important targets for characterisation.  As one example, all four instrument teams for the {\it James Webb Space Telescope\/} have chosen WASP-107b \citep{2017A&A...604A.110A}  as a GTO target. 

The imminent {\it TESS\/} satellite \citep{2016SPIE.9904E..2BR} will perform an all-sky transit survey that is expected to find any hot Jupiters transiting bright stars that the ground-based surveys have missed. In the meantime, several ground-based surveys are making ongoing discoveries including HAT-South \citep[e.g.][]{2018AJ....155...79H}, KELT  \citep[e.g.][]{2018AJ....155..100J}, MASCARA \citep[e.g.][]{2017A&A...606A..73T} and NGTS \citep[e.g.][]{2017arXiv171011099B}.      
Here we report the latest hot-Jupiter discoveries from the WASP-South transit survey, verified with the Euler/CORALIE spectrograph (e.g. \citealt{2013A&A...551A..80T}) and with follow-up photometry from EulerCAM (e.g. \citealt{2012A&A...544A..72L}) and from the robotic TRAPPIST photometer (e.g. \citealt{2013A&A...552A..82G}).

\begin{table}
\caption{Observations\protect\rule[-1.5mm]{0mm}{2mm}}  
\begin{tabular}{lcr}
\hline 
Facility & Date & Notes \\ [0.5mm] \hline
\multicolumn{3}{l}{{\bf WASP-144:}}\\  
WASP-South & 2006 May--2012 Jun & 29\,500 points \\ 
CORALIE  & 2014 Jun--2016 Oct  &   16 RVs \\
TRAPPIST & 2014 Aug 13 & Blue-block\\
TRAPPIST & 2014 Nov 17 & Blue-block\\
TRAPPIST & 2015 May 18 & Blue-block\\
TRAPPIST & 2014 Jun 12 & Blue-block\\
EulerCAM  & 2015 Jun 28 & {\it NGTS} filter \\ 
\multicolumn{3}{l}{{\bf WASP-145:}}\\  
WASP-South & 2008 Jun--2012 Jun & 54\,000 points \\ 
CORALIE  & 2014 Jun--2016 Aug  &   19 RVs \\
TRAPPIST & 2014 Nov 09 & $z$ band \\
EulerCAM  & 2014 Nov 16 & Gunn $z$ filter \\ 
EulerCAM  & 2015 Jul 02 & Gunn $z$ filter \\ 
\multicolumn{3}{l}{{\bf WASP-158:}}\\  
WASP-South & 2008 Jun--2012 Nov & 27\,000 points \\ 
CORALIE  & 2014 Aug--2016 Oct  &   20 RVs \\
TRAPPIST & 2016 Oct 12 & $I+z$ band \\
EulerCAM  & 2016 Nov 03 & {\it NGTS} filter \\ 
\multicolumn{3}{l}{{\bf WASP-159:}}\\  
WASP-South & 2006 Sep--2012 Feb & 43\,800 points \\ 
CORALIE  & 2014 Nov--2017 Mar  &   29 RVs \\
EulerCAM & 2016 Dec  02 & {\it NGTS\/} filter \\ 
\multicolumn{3}{l}{{\bf WASP-162:}}\\  
WASP-South & 2006 May--2012 Jun & 37\,400 points \\ 
CORALIE  & 2014 Apr--2017 May  &   18 RVs \\
TRAPPIST & 2015 Jun 07 & $I+z$ band \\
EulerCAM & 2017 Jan 14 & $I_{c}$ filter \\ 
\multicolumn{3}{l}{{\bf WASP-168:}}\\  
WASP-South & 2006 Oct--2012 Mar & 31\,000 points \\ 
CORALIE  & 2014 Dec--2017 Apr  &   31 RVs \\
TRAPPIST & 2016 Feb 05 & $I+z$ band \\
TRAPPIST & 2016 Sep 16 & $V$ band \\
EulerCAM & 2016 Sep 16 & $I_{c}$ filter \\ 
\multicolumn{3}{l}{{\bf WASP-172:}}\\  
WASP-South & 2006 May--2012 Jun & 35\,000 points \\ 
CORALIE  & 2012 Apr--2017 Jun  &   38 RVs \\
TRAPPIST & 2014 Feb 03 & $I+z$ band \\
TRAPPIST & 2014 Jun 20 & $I+z$ band \\
EulerCAM  & 2015 Jul 03& Gunn $r$ filter \\ 
TRAPPIST & 2015 May 31 & $z^{\prime}$ filter \\ 
\multicolumn{3}{l}{{\bf WASP-173:}}\\  
WASP-South & 2006 May--2011 Nov & 19\,000 points \\ 
CORALIE  & 2015 Sep--2016 Dec  &   18 RVs \\
EulerCAM  & 2016 Sep 21 & Gunn $z$ filter \\ 
TRAPPIST & 2016 Oct 06 & $V$ filter \\
EulerCAM  & 2016 Oct 20 & Gunn $z$ filter \\ 
\end{tabular} 
\end{table}

\section{Observations}
WASP-South is an array of eight cameras that, for the observations reported here,  each consisted of a 200-mm, f/1.8 Canon lenses with a 2k$\times$2k CCD, giving a  $7.8^{\circ}\times7.8^{\circ}$ field.  The cameras are all on the same mount, which rasters a set of fields with a typical 10-min cadence, recording stars in the range  $V$ = 9--13. Processed photometry is accumulated in a central archive where the multi-year dataset on each star is searched for transits (see \citealt{2007MNRAS.380.1230C}).  After vetting of all candidates by eye, the best ones are sent for followup observations with TRAPPIST and  Euler/CORALIE.  The observations for each star are listed in Table~1. All of our methods are similar to those for previous WASP-South discovery papers (e.g.\ \citealt{2014MNRAS.440.1982H}; \citealt{2016A&A...591A..55M}; \citealt{2017MNRAS.465.3693H}), and for this reason the presentation here is relatively concise. 

\section{Spectral analysis}
We report spectral analyses of the host stars made using the CORALIE spectra. Adopting methods as described by \citet{2013MNRAS.428.3164D}, we estimated the effective temperature, $T_{\rm eff}$, from the H$\alpha$ line, and the surface gravity, $\log g$, from Na~{\sc i} D and Mg~{\sc i} b lines.  We also report an indicative spectral type deduced from the $T_{\rm eff}$ estimates.   We report [Fe/H] values determined from equivalent-width measurements of unblended Fe~{\sc i} lines.   The errors that we quote for the abundances take into account the uncertainties in $T_{\rm eff}$ and $\log g$.  The  Fe~{\sc i} lines were also used to estimate the rotation speed, $v \sin i$, after convolving with the CORALIE instrumental resolution ($R$ = 55\,000), and also combining with an estimate of the macroturbulence take from \citet{2014MNRAS.444.3592D}.  Lastly, we report lithium abundance values and corresponding age estimates using \citep{2005A&A...442..615S}, though such estimates are unreliable.   The parameters obtained from the analysis are given in the Tables for each system.  Where available we also list parallax values from the GAIA DR1 \citep{2016A&A...595A...2G} and proper motions from the UCAC5 catalogue \citep{2017AJ....153..166Z}.

\section{Stellar rotational modulations}
The WASP photometry can span months of  a year with observations on each clear night, and so is sensitive to rotational modulations down to the millimag level.  We thus routinely search the photometry of planet hosts using a sine-wave fitting algorithm. We also compute a false alarm probability by repeatedly shuffling the nightly datasets (see \citealt{2011PASP..123..547M}). For most of the planet hosts in this paper we find only upper limits, but for WASP-144 and WASP-173 we found significant modulations.

\section{System parameters}
As we have routinely done for previous planet-discovery papers, we
combine the photometric and radial-velocity data sets for each system
into one Markov-chain Monte-Carlo (MCMC) analysis, using a code originally 
described by \citet{2007MNRAS.375..951C}.  Since the CORALIE spectrograph underwent an upgrade in 2014 November we allow for a radial-velocity offset between the datasets before and after the upgrade (for future reference the RV values are listed in Table~A1, where the time of upgrade is marked by a short line).  The treatment of limb darkening is crucial to fitting transit photometry, and here we have used the 4-parameter, non-linear law of \citet{2000A&A...363.1081C},
interpolating coefficients for the appropriate stellar temperature and
metallicity of each star.

On early MCMC runs we allowed the eccentricity to be a free parameter, and for one of our systems (WASP-162b) we found a highly significant eccentricity. Where, however, the outcome was compatible with a circular orbit, we then computed results with a circular orbit imposed (this gives the most likely set of parameters, as discussed by \citet{2012MNRAS.422.1988A}, essentially feeding in the prior expectation that most hot Jupiters have orbits that have been circularised by tidal forces). 

The list of MCMC parameters for the circular-orbit case is $T_{\rm c}$ (the epoch of mid-transit),  $P$ (the orbital period), $\Delta F$ (the transit depth that would be
observed in the absence of limb-darkening), $T_{14}$ (duration from first to fourth contact), $b$ (the impact parameter) and $K_{\rm 1}$ (the stellar reflex velocity).   The fitted parameters and other values derived from them are listed in a table for each system, along with  1-$\sigma$ errors (though where no eccentricity is found we quote 2-$\sigma$ upper limits).   Red noise in the photometry could mean that the uncertainties are larger than quoted.  For an account of the effects of red noise in datasets similar to those reported here see the  analysis of multiple different transit lightcurves of WASP-36b by \citet{2012AJ....143...81S}.

An additional constraint on the fitting,  continuing our practice from other recent discovery papers, comes from stellar models.  We first run an MCMC analysis to estimate the stellar density (which can be derived from the transit lightcurve independently of a stellar model).   We then use the stellar density and the
spectroscopic effective temperature and metallicity to estimate the most likely stellar mass using the {\sc bagemass} code described in  \citet{2015A&A...575A..36M}, which is based on  the {\sc garstec} stellar evolution code
\citep{2008Ap&SS.316...99W}.  We then use a resulting stellar-mass estimate and its  error as a Gaussian-prior input to the final MCMC analysis.  

\begin{figure}
\resizebox{\hsize}{!}{\includegraphics{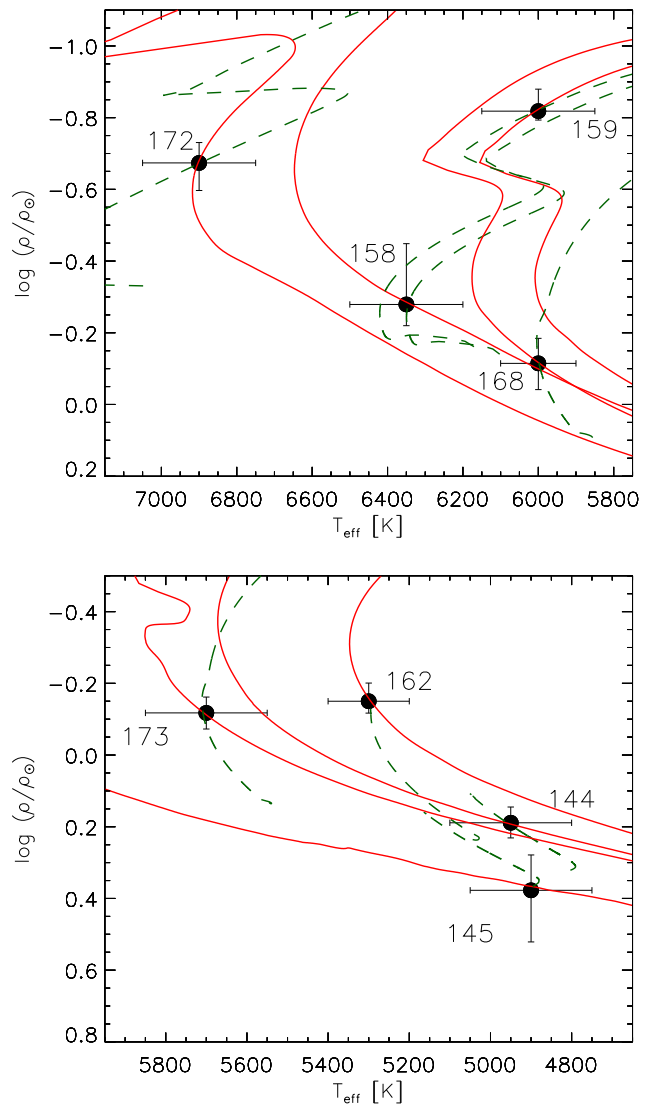}}
\caption{The host star's effective temperature (T$_{\rm eff}$) versus density (each symbol being labelled by the WASP planet number).  We show  
best-fit evolution tracks (dashed lines) and isochrones (solid lines) 
for the masses, ages and [Fe/H] values  listed in Table~\ref{ResultsTable}.
\label{trho_plot}
}
\end{figure}

The masses and ages of the stars derived from {\sc bagemass} are given in
Table~\ref{ResultsTable}.  The best-fit stellar evolution tracks and
isochrones are shown in Fig.~\ref{trho_plot}. 

\begin{table*}
 \caption{Mass and age estimates for the host stars, derived from the {\sc bagemass} code with {\sc garstec} stellar models and assuming $\alpha_{\rm
MLT}=1.78$. Columns 2, 3 and 4 give the maximum-likelihood estimates of the
age, mass, and initial metallicity, respectively. Column 5 is the
chi-squared statistic of the fit for the parameter values in columns 2, 3, and
4.  Columns 6 and 7 give the mean and standard deviation of their posterior
distributions. The systematic errors on the mass and age due to uncertainties
in the mixing length and helium abundance are given in columns 8 to 11.
\label{ResultsTable}}
 \begin{tabular}{@{}lrrrrrrrrrr}
\hline
\hline
Star &
  \multicolumn{1}{c}{$\tau_{\rm iso, b}$ [Gyr]} &
  \multicolumn{1}{c}{$M_{\rm b}$[$M_{\odot}$]} &
  \multicolumn{1}{c}{$\mathrm{[Fe/H]}_\mathrm{i, b}$} &
  \multicolumn{1}{c}{$\chi^2$}&
  \multicolumn{1}{c}{$\langle \tau_{\rm iso} \rangle$ [Gyr]}  &
  \multicolumn{1}{c}{$\langle M_{\star} \rangle$ [$M_{\odot}$]} &
  \multicolumn{1}{c}{$\sigma_{\tau, Y}$}  &
  \multicolumn{1}{c}{$\sigma_{\tau,\alpha}$} &
  \multicolumn{1}{c}{$\sigma_{M, Y}$}  &
  \multicolumn{1}{c}{$\sigma_{M,\alpha}$}  \\
\hline
 \noalign{\smallskip}
WASP-144         &   9.5 &   0.84 &$ +0.238 $&  0.01 &$ 8.71 \pm  4.12 $&$  0.844 \pm 0.046 $&$  1.96 $&$  3.96 $&$ -0.040 $&$ -0.026 $ \\
WASP-145         &   0.0 &   0.79 &$ -0.048 $&  0.07 &$ 6.99 \pm  4.40 $&$  0.763 \pm 0.040 $&$  0.12 $&$  0.02 $&$ -0.034 $&$ -0.006 $ \\
WASP-158         &   1.5 &   1.32 &$ +0.252 $&  0.01 &$ 1.93 \pm  0.93 $&$  1.339 \pm 0.092 $&$  0.01 $&$  0.75 $&$ -0.035 $&$ -0.030 $ \\
WASP-159         &   4.1 &   1.33 &$ +0.227 $&  0.03 &$ 3.40 \pm  0.95 $&$  1.431 \pm 0.118 $&$ -0.08 $&$  1.04 $&$ -0.036 $&$ -0.103 $ \\
WASP-162         &  13.7 &   0.94 &$ +0.381 $&  0.00 &$12.97 \pm  2.35 $&$  0.953 \pm 0.041 $&$  0.84 $&$  3.10 $&$ -0.044 $&$ -0.034 $ \\
WASP-168         &   3.9 &   1.08 &$ +0.046 $&  0.01 &$ 3.96 \pm  1.77 $&$  1.073 \pm 0.053 $&$  0.35 $&$  1.65 $&$ -0.043 $&$ -0.034 $ \\
WASP-172         &   1.7 &   1.48 &$ -0.102 $&  0.04 &$ 1.79 \pm  0.28 $&$  1.472 \pm 0.067 $&$  0.00 $&$  0.03 $&$ -0.050 $&$ -0.006 $ \\
WASP-173         &   6.6 &   1.03 &$ +0.225 $&  0.00 &$ 6.78 \pm  2.93 $&$  1.035 \pm 0.072 $&$  0.51 $&$  2.35 $&$ -0.044 $&$ -0.037 $ \\
 \noalign{\smallskip}
\hline
\end{tabular}   
\end{table*}

\begin{figure}
\hspace*{2mm}\includegraphics[width=8.5cm]{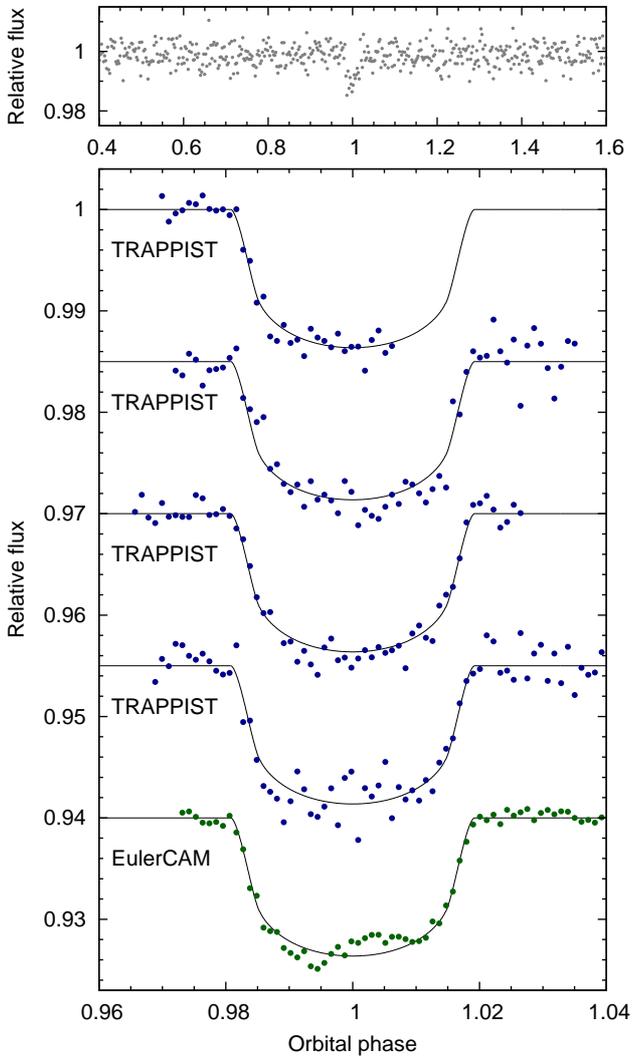}\\ [-2mm]
\caption{WASP-144b discovery photometry: (Top) The WASP data folded on the 
transit period. (Second panel) The binned WASP data with (offset) the
follow-up transit lightcurves (ordered from the top as in Table~1) together with the fitted MCMC model.}
\end{figure}

\begin{figure}
\hspace*{2mm}\includegraphics[width=8.5cm]{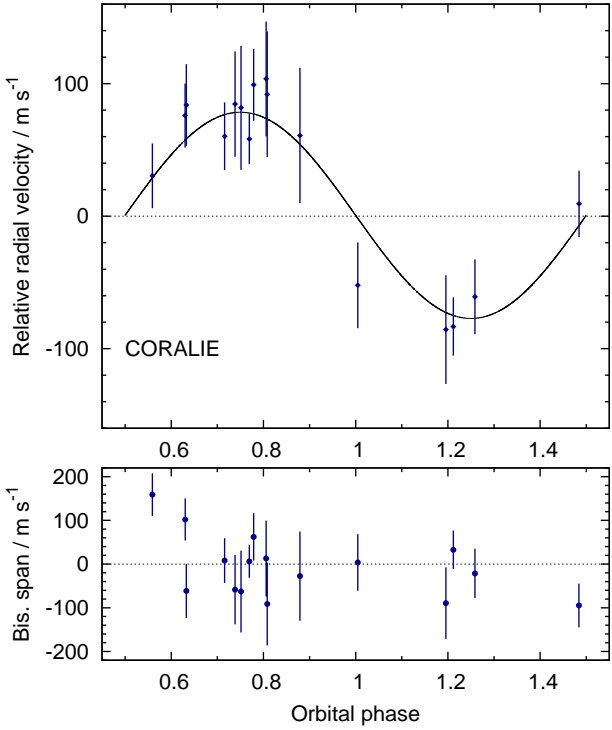}\\ [-2mm]
\caption{WASP-144b radial velocities and fitted model (top) along with (bottom) the bisector spans; the absence of any correlation with radial velocity is a check against transit mimics.}
\end{figure}

\begin{table}
\caption{System parameters for WASP-144.}  
\begin{tabular}{lc}
\multicolumn{2}{l}{1SWASP\,J212303.08--400253.6}\\
\multicolumn{2}{l}{2MASS\,21230309--4002537}\\
\multicolumn{2}{l}{GAIA RA\,=\,21$^{\rm h}$23$^{\rm m}$03.09$^{\rm s}$, 
Dec\,=\,--40$^{\circ}$02$^{'}$54.4$^{''}$ (J2000)}\\
\multicolumn{2}{l}{$V$ mag = 12.9}  \\ 
\multicolumn{2}{l}{Rotational modulation: $P$ = 21 $\pm$ 1 d, 4--8 mmag}\\
\multicolumn{2}{l}{UCAC5 pm (RA) --3.4\,$\pm$\,1.0 (Dec) --41.7\,$\pm$\,1.0 mas/yr}\\
\hline
\multicolumn{2}{l}{Stellar parameters from spectroscopic analysis.\rule[-1.5mm]{0mm}{2mm}} \\ \hline 
Spectral type & K2V \\
$T_{\rm eff}$ (K)  & 4950  $\pm$ 150  \\
$\log g$      & 4.5 $\pm$ 0.2    \\
$v\,\sin i$ (km\,s$^{-1}$)     &   1.9 $\pm$ 1.2     \\
{[Fe/H]}   &  +0.18 $\pm$ 0.14     \\
log A(Li)  &    $<$    0.6    \\
Age (Lithium) [Gy]  &  $>$   0.5      \\ \hline 
\multicolumn{2}{l}{Parameters from MCMC analysis.\rule[-1.5mm]{0mm}{3mm}} \\
\hline 
$P$ (d) & 2.2783152 $\pm$ 0.0000013 \\
$T_{\rm c}$ (HJD)\,(UTC) & 245\,7157.27493 $\pm$ 0.00015 \\
$T_{\rm 14}$ (d) & 0.0814 $\pm$ 0.0007 \\
$T_{\rm 12}=T_{\rm 34}$ (d) & 0.0104 $\pm$ 0.0009 \\
$\Delta F=R_{\rm P}^{2}$/R$_{*}^{2}$ & 0.01165 $\pm$ 0.00028 \\
$b$ & 0.45 $\pm$ 0.07 \\
$i$ ($^\circ$)  & 86.9 $\pm$ 0.5 \\
$K_{\rm 1}$ (km s$^{-1}$) & 0.078 $\pm$ 0.011 \\
$\gamma$ (km s$^{-1}$)  & 16.105 $\pm$ 0.008 \\
$e$ & 0 (adopted) ($<$\,0.30 at 2$\sigma$) \\ 
$a/R_{\rm *}$  & 8.39 $\pm$ 0.23 \\ 
$M_{\rm *}$ (M$_{\rm \odot}$) & 0.81 $\pm$ 0.04 \\
$R_{\rm *}$ (R$_{\rm \odot}$) & 0.81 $\pm$ 0.04 \\
$\log g_{*}$ (cgs) & 4.53 $\pm$ 0.03 \\
$\rho_{\rm *}$ ($\rho_{\rm \odot}$) & 1.54 $\pm$ 0.16\\
$T_{\rm eff}$ (K) & 5200 $\pm$ 140 \\
$M_{\rm P}$ (M$_{\rm Jup}$) & 0.44 $\pm$ 0.06 \\
$R_{\rm P}$ (R$_{\rm Jup}$) & 0.85 $\pm$ 0.05 \\
$\log g_{\rm P}$ (cgs) & 3.15 $\pm$ 0.06 \\
$\rho_{\rm P}$ ($\rho_{\rm J}$) & 0.72 $\pm$ 0.15 \\
$a$ (AU)  & 0.0316  $\pm$ 0.0005 \\
$T_{\rm P, A=0}$ (K) & 1260 $\pm$ 40 \\ [0.5mm] \hline 
\multicolumn{2}{l}{Errors are 1$\sigma$; Limb-darkening coefficients were:}\\
\multicolumn{2}{l}{{\small $r$ band: a1 = 0.734 a2 = --0.714, a3 = 1.399, 
a4 = --0.614}}\\ \hline
\end{tabular} 
\end{table}

\begin{figure}
\hspace*{0mm}\includegraphics[width=9cm]{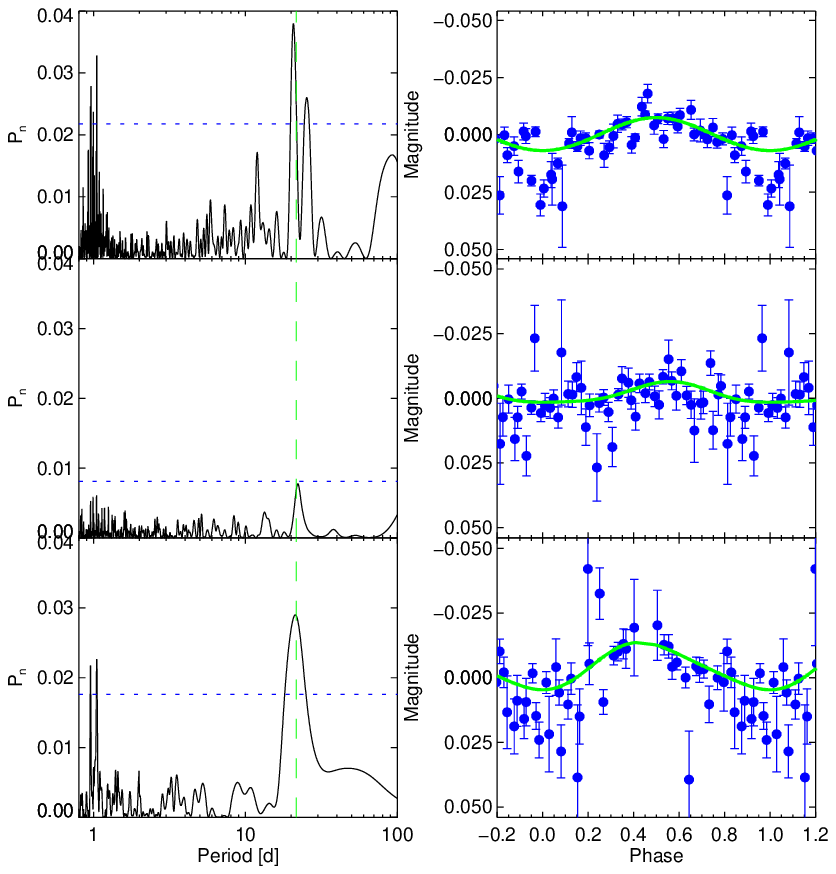}\\ [-2mm]
\caption{WASP-144 rotational modulation. The left-hand panels show periodograms for independent WASP-South data sets from 2006, 2011 and 2012. The blue dotted line is a false-alarm probability of 0.001.  The right-hand panels show the data for each season folded on the 21-d period; the green line is a harmonic-series fit.}
\end{figure}

\section{WASP-144}
The discovery photometry and radial-velocity data for WASP-144 are shown in Figs.~2 \&\ 3. We also show the bisector spans, which are a check for transit mimics. The system parameters are listed in Table~3.  WASP-144 is a relatively faint, $V$ = 12.9, K2 star with a metallicity of [Fe/H] = +0.18 $\pm$ 0.14.    Evolutionary tracks suggest that it could be 8 $\pm$ 4 Gy old (Table~\ref{ResultsTable}). 

The WASP data show a possible rotational modulation at a period of 21 $\pm$ 1 d with an amplitude of 4--8 mmag (see Fig.~4).    This is seen independently in data from the seasons 2006 (7 mmag amplitude; false-alarm probability $<$0.1\%), 2011 (4-mmag amplitude; FAP 6\%) and 2012 (8-mmag amplitude, FAP $<$0.1\%).  The rotational period and fitted radius (0.81 $\pm$ 0.04 R$_{\odot}$) imply a surface rotational velocity of 1.96 $\pm$ 0.13 km s$^{-1}$, which compares with an observed \vsini\ of 1.9 $\pm$ 1.2 km s$^{-1}$.    Thus this is likely to be an aligned system (with the stellar spin axis perpendicular to us). 

The followup photometry consists of four TRAPPIST transits and one from 
EulerCAM.  The latter was observed in poor conditions with variable seeing, leading to systematic features in the lightcurve that we do not think are astrophysically real.  The MCMC process down-weighted this lightcurve in the fitting. 

WASP-144b has a mass of 0.44 M$_{\rm Jup}$ and a radius of 0.85 R$_{\rm Jup}$ and is  in a 2.27-d orbit.     The radius of  0.85 R$_{\rm Jup}$ is among the lowest found for hot-Jupiter planets.  Comparable planets are WASP-60b (0.50 M$_{\rm Jup}$; 0.90 R$_{\rm Jup}$; \citealt{2013A&A...549A.134H}) and Kepler-41b (0.55 M$_{\rm Jup}$; 0.89 R$_{\rm Jup}$; \citealt{2011A&A...536A..70S}. Both of those have G-star hosts whereas WASP-144 is a K2 star.

\begin{table}
\caption{System parameters for WASP-145.}  
\begin{tabular}{lc}
\multicolumn{2}{l}{1SWASP\,J212900.65--585008.4}\\
\multicolumn{2}{l}{2MASS\,212900.65--585008.4}\\
\multicolumn{2}{l}{GAIA RA\,=\,21$^{\rm h}$29$^{\rm m}$00.90$^{\rm s}$, 
Dec\,=\,--58$^{\circ}$50$^{'}$10.1$^{''}$ (J2000)}\\
\multicolumn{2}{l}{$V$ mag = 11.5}  \\ 
\multicolumn{2}{l}{Rotational modulation\ \ \ $<$ 2 mmag (95\%)}\\
\multicolumn{2}{l}{UCAC4 pm (RA) 102.6\,$\pm$\,1.2 (Dec) 4.9\,$\pm$\,3.4 mas/yr}\\
\hline
\multicolumn{2}{l}{Stellar parameters from spectroscopic analysis.\rule[-1.5mm]{0mm}{2mm}} \\ \hline 
Spectral type & K2V \\
$T_{\rm eff}$ (K)  & 4900  $\pm$ 150  \\
$\log g$      & 4.6 $\pm$ 0.2    \\
$v\,\sin i$ (km\,s$^{-1}$)     &    2.1 $\pm$ 1.1     \\
{[Fe/H]}   &   $-$0.04 $\pm$ 0.10     \\
log A(Li)  &    $<$ 0.5      \\
Age (Lithium) [Gy]  &  $>$  0.5       \\ \hline 
\multicolumn{2}{l}{Parameters from MCMC analysis.\rule[-1.5mm]{0mm}{3mm}} \\
\hline 
$P$ (d) & 1.7690381 $\pm$ 0.0000008 \\
$T_{\rm c}$ (HJD)\,(UTC) & 245\,6844.16526 $\pm$ 0.00026 \\
$T_{\rm 14}$ (d) & 0.0407 $\pm$ 0.0016 \\
$T_{\rm 12}=T_{\rm 34}$ (d) & undefined \\
$\Delta F=R_{\rm P}^{2}$/R$_{*}^{2}$ & 0.0116 $\pm$ 0.0026 \\
$b$ & 0.97 $\pm$ 0.09 \\
$i$ ($^\circ$)  & 83.3 $\pm$ 1.3 \\
$K_{\rm 1}$ (km s$^{-1}$) & 0.178 $\pm$ 0.006 \\
$\gamma$ (km s$^{-1}$)  & 3.345 $\pm$ 0.005 \\
$e$ & 0 (adopted) ($<$\,0.06 at 2$\sigma$) \\ 
$a/R_{\rm *}$  & 8.74 $\pm$ 0.52\\ 
$M_{\rm *}$ (M$_{\rm \odot}$) & 0.76 $\pm$ 0.04 \\
$R_{\rm *}$ (R$_{\rm \odot}$) & 0.68 $\pm$ 0.07 \\
$\log g_{*}$ (cgs) & 4.65 $\pm$ 0.10 \\
$\rho_{\rm *}$ ($\rho_{\rm \odot}$) & 2.38 $\pm$ 0.93\\
$T_{\rm eff}$ (K) & 4900 $\pm$ 150 \\
$M_{\rm P}$ (M$_{\rm Jup}$) & 0.89 $\pm$ 0.04 \\
$R_{\rm P}$ (R$_{\rm Jup}$) & 0.9 $\pm$ 0.4 \\
$\log g_{\rm P}$ (cgs) & 3.4 $\pm$ 0.4 \\
$\rho_{\rm P}$ ($\rho_{\rm J}$) & 1.2 $\pm$ 1.0 \\
$a$ (AU)  & 0.0261  $\pm$ 0.0005 \\
$T_{\rm P, A=0}$ (K) & 1200 $\pm$ 60 \\ [0.5mm] \hline 
\multicolumn{2}{l}{Errors are 1$\sigma$; Limb-darkening coefficients were:}\\
\multicolumn{2}{l}{{\small $r$ band: a1 = 0.703 a2 = --0.734, a3 = 1.472, 
a4 = --0.630}}\\ 
\multicolumn{2}{l}{{\small $z$ band: a1 =   0.765 , a2 = --0.800 , a3 = 1.258, a4 = --0.530}}\\ \hline
\end{tabular} 
\end{table}

\begin{figure}
\hspace*{2mm}\includegraphics[width=8.5cm]{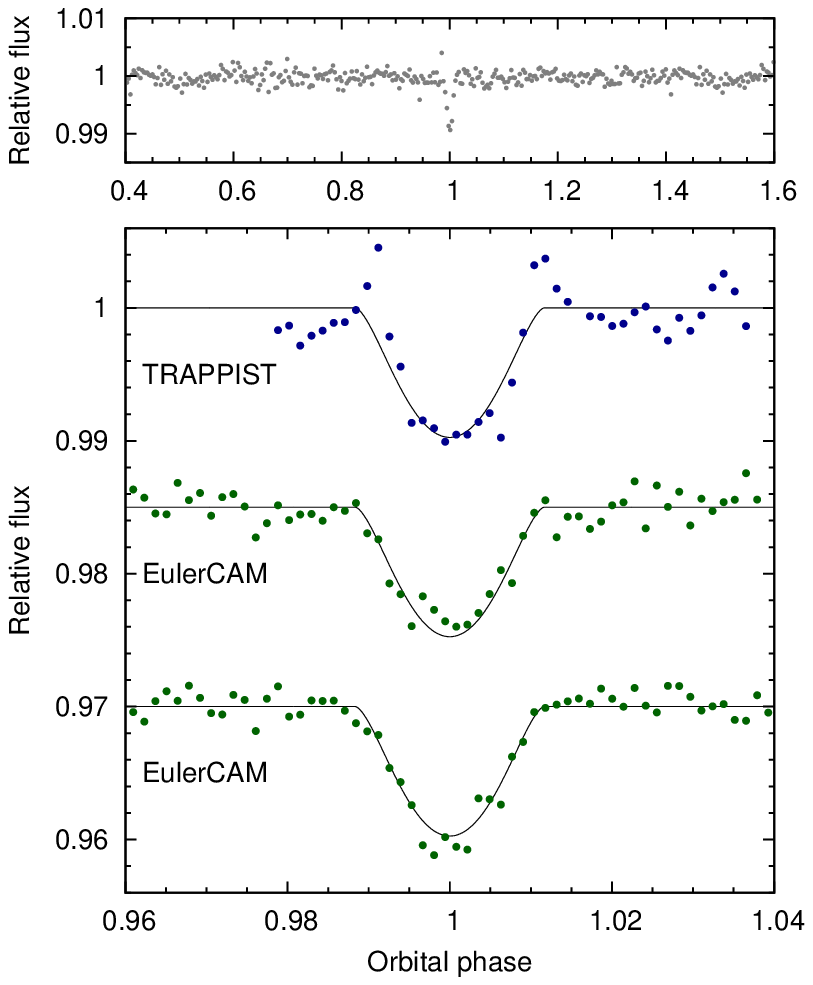}\\ [-2mm]
\caption{WASP-145b discovery photometry, as for Fig.~2.}
\end{figure}

\begin{figure}
\hspace*{2mm}\includegraphics[width=8.5cm]{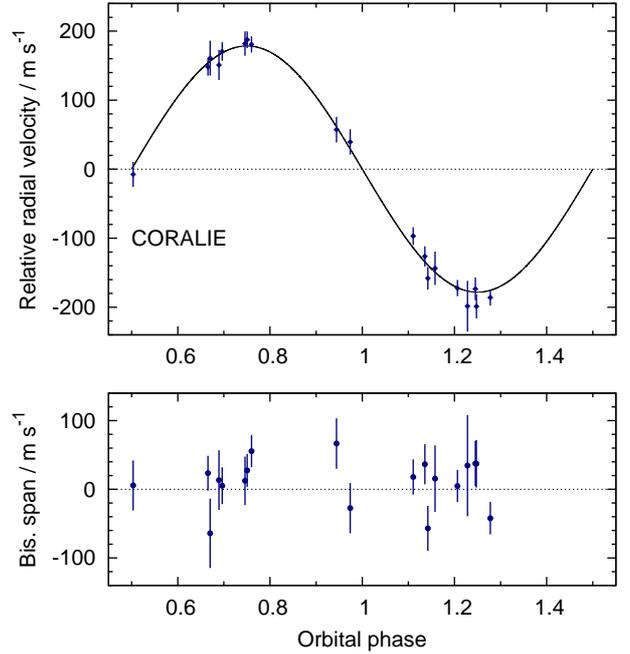}\\ [-2mm]
\caption{WASP-145b radial velocities and bisector spans, as for Fig.~3.}
\end{figure}

\section{WASP-145}
WASP-145A is a $V$ = 11.5, K2 star, with solar metallicity ([Fe/H] = +0.04 $\pm$ 0.10) and an  estimated age of 7 $\pm$ 4 Gy (Figs.~5 \&\ 6; Table~4).

WASP-145A has a companion star, WASP-145B, separated by 5.14 $\pm$ 0.01 arcsecs with a position angle of --5.04 $\pm$ 0.13 degrees, and fainter by 1.407 $\pm$ 0.009 mag (as measured by focused EulerCAM images in a $z^{\prime}$-band filter). It is likely that this star is physically associated with WASP-145A, but this is not certain.   The later EulerCAM lightcurve (July 2015) was extracted using an aperture including both stars and this lightcurve was corrected for dilution in the analysis (the other follow-up lightcurves used a smaller aperture containing only the host star). 

The planet WASP-145Ab has a 1.77-d orbit and a grazing transit ($b$ = 0.97 $\pm$ 0.09), which means that 2$^{\rm nd}$ and 3$^{\rm rd}$ contact are not discernable and thus that the planetary radius is not well constrained. We estimate the mass at 0.89 $\pm$ 0.04 M$_{\rm Jup}$ and the radius at 0.9 $\pm$ 0.4 R$_{\rm Jup}$.  If the radius were at the lower end of that range it would be abnormally low for a hot Jupiter.  The transit depth of 1.1\%\ is typical for a  hot Jupiter owing to a relatively small stellar radius of 0.68 $\pm$ 0.07 R$_{\odot}$.

\begin{table}
\caption{System parameters for WASP-158.}  
\begin{tabular}{lc}
\multicolumn{2}{l}{1SWASP\,J001635.09--105834.9}\\
\multicolumn{2}{l}{2MASS\,001635.09--105834.9}\\
\multicolumn{2}{l}{GAIA RA\,=\,00$^{\rm h}$16$^{\rm m}$35.12$^{\rm s}$, 
Dec\,=\,--10$^{\circ}$58$^{'}$35.1$^{''}$ (J2000)}\\
\multicolumn{2}{l}{$V$ mag = 12.1}  \\ 
\multicolumn{2}{l}{Rotational modulation\ \ \ $<$ 1.5 mmag  (95\%)}\\
\multicolumn{2}{l}{UCAC5 pm (RA) 2.4\,$\pm$\,1.2 (Dec) 0.4\,$\pm$\,1.2 mas/yr}\\
\multicolumn{2}{l}{GAIA DR1 parallax: 0.88 $\pm$ 0.57 mas}\\
\hline
\multicolumn{2}{l}{Stellar parameters from spectroscopic analysis.\rule[-1.5mm]{0mm}{2mm}} \\ \hline 
Spectral type & F6V \\
$T_{\rm eff}$ (K)  & 6350  $\pm$ 150  \\
$\log g$      & 4.5 $\pm$ 0.2    \\
$v\,\sin i$ (km\,s$^{-1}$)     &    9.3 $\pm$ 1.3     \\
{[Fe/H]}   &  +0.24 $\pm$ 0.10     \\
log A(Li)  &    $<$   1.7   \\
Age (Lithium) [Gy]  & (close to Lithium gap)      \\ \hline 
\multicolumn{2}{l}{Parameters from MCMC analysis.\rule[-1.5mm]{0mm}{3mm}} \\
\hline 
$P$ (d) & 3.656333 $\pm$ 0.000004 \\
$T_{\rm c}$ (HJD)\,(UTC) & 245\,7619.9195 $\pm$ 0.0010 \\
$T_{\rm 14}$ (d) & 0.1490 $\pm$ 0.0037 \\
$T_{\rm 12}=T_{\rm 34}$ (d) & 0.014 $\pm$ 0.002 \\
$\Delta F=R_{\rm P}^{2}$/R$_{*}^{2}$ & 0.0063 $\pm$ 0.0005 \\
$b$ & 0.32 $\pm$ 0.23 \\
$i$ ($^\circ$)  & 87.7 $\pm$ 1.5 \\
$K_{\rm 1}$ (km s$^{-1}$) & 0.295 $\pm$ 0.015 \\
$\gamma$ (km s$^{-1}$)  & 24.197 $\pm$ 0.012 \\
$e$ & 0 (adopted) ($<$\,0.16 at 2$\sigma$) \\ 
$a/R_{\rm *}$  & $8.0^{+0.4}_{-1.0}$\\ 
$M_{\rm *}$ (M$_{\rm \odot}$) & 1.38 $\pm$ 0.14 \\
$R_{\rm *}$ (R$_{\rm \odot}$) & 1.39 $\pm$ 0.18 \\
$\log g_{*}$ (cgs) & 4.30 $^{+0.05}_{-0.11}$ \\
$\rho_{\rm *}$ ($\rho_{\rm \odot}$) & 0.53 $^{+0.08}_{-0.17}$\\
$T_{\rm eff}$ (K) & 6350 $\pm$ 150 \\
$M_{\rm P}$ (M$_{\rm Jup}$) & 2.79 $\pm$ 0.23 \\
$R_{\rm P}$ (R$_{\rm Jup}$) & 1.07 $\pm$ 0.15 \\
$\log g_{\rm P}$ (cgs) & 3.75 $^{+0.06}_{-0.14}$  \\
$\rho_{\rm P}$ ($\rho_{\rm J}$) & 2.3 $^{+0.5}_{-0.9}$ \\
$a$ (AU)  & 0.0517  $\pm$ 0.0018 \\
$T_{\rm P, A=0}$ (K) & 1590 $\pm$ 80 \\ [0.5mm] \hline 
\multicolumn{2}{l}{Errors are 1$\sigma$; Limb-darkening coefficients were:}\\
\multicolumn{2}{l}{{\small $r$ band: a1 = 0.568, a2 = 0.137, a3 = 0.145, 
a4 = --0.136}}\\ 
\multicolumn{2}{l}{{\small $z$ band: a1 =   0.658 , a2 = --0.252 , a3 = 0.422, a4 = --0.226}}\\ \hline
\end{tabular} 
\end{table}

\section{WASP-158}
WASP-158 is a $V$ = 12.1, F6 star with a metallicity of [Fe/H] = +0.24 $\pm$ 0.15. It is compatible with being an un-evolved main-sequence star with an age estimate of 1.9 $\pm$ 0.9 Gy.   

The planet, WASP-158b, is relatively massive at 2.8 M$_{\rm Jup}$ with a radius of 1.1 R$_{\rm Jup}$, and has a 3.66-d orbit (Fig.~7; Table~5). WASP-158b is thus very similar to  WASP-38b (2.7 M$_{\rm Jup}$; 1.1 R$_{\rm Jup}$, in a 6.9-d orbit around an F8 star;  \citealt{2011A&A...525A..54B}) and WASP-99b (2.8 M$_{\rm Jup}$; 1.1 R$_{\rm Jup}$, in a 5.8-d orbit around an F8 star; \citealt{2014MNRAS.440.1982H}).

\begin{figure}
\hspace*{2mm}\includegraphics[width=8.5cm]{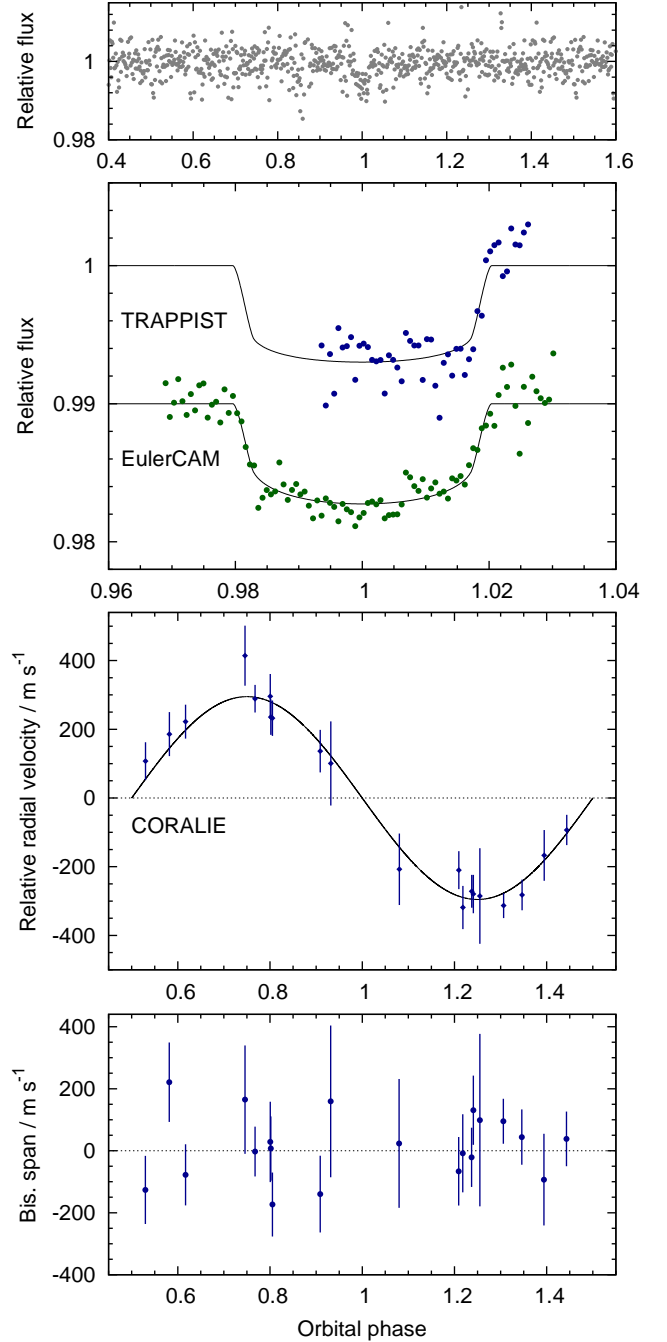}\\ [-2mm]
\caption{WASP-158b discovery data, as for Figs.~2 \&\ 3.}
\end{figure}

\clearpage

\begin{table}
\caption{System parameters for WASP-159.}  
\begin{tabular}{lc}
\multicolumn{2}{l}{1SWASP\,J043232.73--385805.8}\\
\multicolumn{2}{l}{2MASS\,04323274--3858060}\\
\multicolumn{2}{l}{GAIA RA\,=\,04$^{\rm h}$32$^{\rm m}$32.76$^{\rm s}$, 
Dec\,=\,--38$^{\circ}$58$^{'}$06.0$^{''}$ (J2000)}\\
\multicolumn{2}{l}{$V$ mag = 12.8}  \\ 
\multicolumn{2}{l}{Rotational modulation\ \ \ $<$ 1.5 mmag  (95\%)}\\
\multicolumn{2}{l}{UCAC5 pm (RA) --0.9\,$\pm$\,1.0 (Dec) 5.2\,$\pm$\,1.0 mas/yr}\\
\multicolumn{2}{l}{GAIA parallax: 1.08  $\pm$ 0.24 mas}\\
\hline
\multicolumn{2}{l}{Stellar parameters from spectroscopic analysis.\rule[-1.5mm]{0mm}{2mm}} \\ \hline 
Spectral type & F9 \\
$T_{\rm eff}$ (K)  & 6000  $\pm$ 150  \\
$\log g$      & 4.0 $\pm$ 0.1    \\
$v\,\sin i$ (km\,s$^{-1}$)     &   5.7 $\pm$ 0.4     \\
{[Fe/H]}   &  +0.22 $\pm$ 0.12     \\
log A(Li)  &   2.15 $\pm$ 0.12       \\
Age (Lithium) [Gy]  &   2 to 8        \\ \hline 
\multicolumn{2}{l}{Parameters from MCMC analysis.\rule[-1.5mm]{0mm}{3mm}} \\
\hline 
$P$ (d) & 3.840401 $\pm$ 0.000007 \\
$T_{\rm c}$ (HJD)\,(UTC) & 245\,7668.0849 $\pm$ 0.0009 \\
$T_{\rm 14}$ (d) & 0.2328 $\pm$ 0.0021 \\
$T_{\rm 12}=T_{\rm 34}$ (d) & 0.0152 $\pm$ 0.0016 \\
$\Delta F=R_{\rm P}^{2}$/R$_{*}^{2}$ & 0.00453 $\pm$ 0.00018 \\
$b$ & 0.18 $\pm$ 0.15  \\
$i$ ($^\circ$)  & 88.1 $\pm$ 1.4 \\
$K_{\rm 1}$ (km s$^{-1}$) & 0.057 $\pm$ 0.008 \\
$\gamma$ (km s$^{-1}$)  & 35.160 $\pm$ 0.006 \\
$e$ & 0 (adopted) ($<$\,0.18 at 2$\sigma$) \\ 
$a/R_{\rm *}$  & $5.44^{+0.15}_{-0.29}$ \\
$M_{\rm *}$ (M$_{\rm \odot}$) & 1.41 $\pm$ 0.12 \\
$R_{\rm *}$ (R$_{\rm \odot}$) & 2.11 $\pm$ 0.10 \\
$\log g_{*}$ (cgs) & 3.94 $\pm$ 0.04  \\
$\rho_{\rm *}$ ($\rho_{\rm \odot}$) & 0.15 $\pm$ 0.02\\
$T_{\rm eff}$ (K) & 6120 $\pm$ 140 \\
$M_{\rm P}$ (M$_{\rm Jup}$) & 0.55 $\pm$ 0.08 \\
$R_{\rm P}$ (R$_{\rm Jup}$) & 1.38 $\pm$ 0.09 \\
$\log g_{\rm P}$ (cgs) & 2.82 $\pm$ 0.07 \\
$\rho_{\rm P}$ ($\rho_{\rm J}$) & 0.21 $\pm$ 0.04 \\
$a$ (AU)  & 0.0538  $\pm$ 0.0015 \\
$T_{\rm P, A=0}$ (K) & 1850 $\pm$ 50 \\ [0.5mm] \hline 
\multicolumn{2}{l}{Errors are 1$\sigma$; Limb-darkening coefficients were:}\\
\multicolumn{2}{l}{{\small $r$ band: a1 = 0.595, a2 = 0.011, a3 = 0.345, 
a4 = --0.220}}\\ \hline 
\end{tabular} 
\end{table}

\section{WASP-159}
WASP-159  (Fig.~8; Table~6) is a fainter, $V$ = 12.9, F9 star with a metallicity of [Fe/H] = +0.22 $\pm$ 0.12.  It appears to be evolving off the main sequence with a radius of 2.1 $\pm$ 0.1 R$_{\rm \odot}$ and an age estimate of 3.4 $\pm$ 1.0 Gy (Fig.~1; Table~2). 

The expanded stellar radius means that the transit depth, at 0.45\%, is relatively small for a ground-based discovery, which then implies a fairly bloated planet (1.4 R$_{\rm Jup}$ and 0.55 M$_{\rm Jup}$). This is a well-populated region of a hot-Jupiter mass--radius plot, no doubt in part because bloated planets are easiest to find in ground-based transit surveys. 

\begin{figure}
\hspace*{2mm}\includegraphics[width=8.5cm]{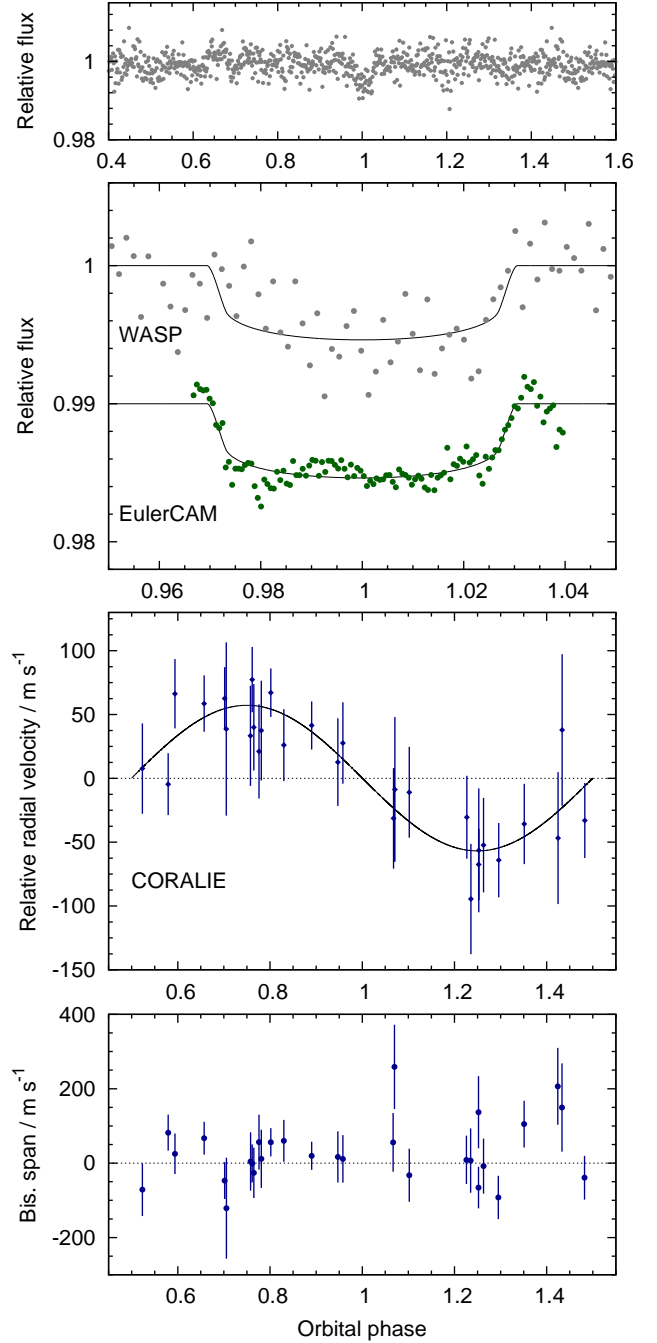}\\ [-2mm]
\caption{WASP-159b discovery data as for Figs.~2 \&\ 3.}
\end{figure}

\begin{table}
\caption{System parameters for WASP-162.}  
\begin{tabular}{lc}
\multicolumn{2}{l}{1SWASP\,J111310.29--173928.1}\\
\multicolumn{2}{l}{2MASS\,11131028--1739280}\\
\multicolumn{2}{l}{GAIA RA\,=\,11$^{\rm h}$13$^{\rm m}$10.30$^{\rm s}$, 
Dec\,=\,--17$^{\circ}$39$^{'}$28.0$^{''}$ (J2000)}\\
\multicolumn{2}{l}{$V$ mag = 12.2}  \\ 
\multicolumn{2}{l}{Rotational modulation\ \ \ $<$ 1 mmag  (95\%)}\\
\multicolumn{2}{l}{UCAC5 pm (RA) 6.2\,$\pm$\,1.2 (Dec) --8.1\,$\pm$\,1.2 mas/yr}\\
\hline
\multicolumn{2}{l}{Stellar parameters from spectroscopic analysis.\rule[-1.5mm]{0mm}{2mm}} \\ \hline 
Spectral type & K0 \\
$T_{\rm eff}$ (K)  & 5300  $\pm$ 100  \\
$\log g$      & 4.5 $\pm$ 0.1    \\
$v\,\sin i$ (km\,s$^{-1}$)     &   1.0 $\pm$ 0.8     \\
{[Fe/H]}   &  +0.28 $\pm$ 0.13     \\
log A(Li)  &    $<$ 0.7     \\
Age (Lithium) [Gy]  &  $>$ 1         \\ \hline 
\multicolumn{2}{l}{Parameters from MCMC analysis.\rule[-1.5mm]{0mm}{3mm}} \\
\hline 
$P$ (d) & 9.62468  $\pm$ 0.00001 \\
$T_{\rm c}$ (HJD)\,(UTC) & 245\,7701.3816 $\pm$ 0.0006 \\
$T_{\rm 14}$ (d) & 0.1774 $\pm$ 0.0015 \\
$T_{\rm 12}=T_{\rm 34}$ (d) & 0.016 $\pm$ 0.001 \\
$\Delta F=R_{\rm P}^{2}$/R$_{*}^{2}$ & 0.0087 $\pm$ 0.0003 \\
$b$ & 0.18 $\pm$ 0.14 \\
$i$ ($^\circ$)  & 89.3 $\pm$ 0.5 \\
$K_{\rm 1}$ (km s$^{-1}$) & 0.507 $\pm$ 0.008 \\
$\gamma$ (km s$^{-1}$)  & 16.824 $\pm$ 0.004 \\
$e$ & 0.434 $\pm$ 0.005 \\ 
$\omega$ (deg) & $-1.9$ $\pm$ 2.2 \\ 
$a/R_{\rm *}$  & $17.0^{+0.4}_{-0.6}$\\ 
$M_{\rm *}$ (M$_{\rm \odot}$) & 0.95 $\pm$ 0.04 \\
$R_{\rm *}$ (R$_{\rm \odot}$) & 1.11 $\pm$ 0.05 \\
$\log g_{*}$ (cgs) & 4.33 $\pm$ 0.03  \\
$\rho_{\rm *}$ ($\rho_{\rm \odot}$) & 0.71 $\pm$ 0.07\\
$T_{\rm eff}$ (K) & 5300 $\pm$ 100 \\
$M_{\rm P}$ (M$_{\rm Jup}$) & 5.2 $\pm$ 0.2 \\
$R_{\rm P}$ (R$_{\rm Jup}$) & 1.00 $\pm$ 0.05 \\
$\log g_{\rm P}$ (cgs) & 4.33 $\pm$ 0.03 \\
$\rho_{\rm P}$ ($\rho_{\rm J}$) & 5.2  $\pm$ 0.6 \\
$a$ (AU)  & 0.0871   $\pm$ 0.0013 \\
$T_{\rm P, A=0}$ (K) & 910 $\pm$ 20 \\ [0.5mm] \hline 
\multicolumn{2}{l}{Errors are 1$\sigma$; Limb-darkening coefficients were:}\\
\multicolumn{2}{l}{{\small $r$ band: a1 = 0.750, a2 = --0.724, a3 = 1.393, 
a4 = --0.614}}\\ 
\multicolumn{2}{l}{{\small $z$ band: a1 =   0.826 , a2 = --0.863 , a3 = 1.266, a4 = --0.540}}\\ 
\multicolumn{2}{l}{{\small $I$ band: a1 =   0.827 , a2 = --0.881 , a3 = 1.376, a4 = --0.598}}\\ \hline
\end{tabular} 
\end{table}

\section{WASP-162}
WASP-162 is a $V$ = 12.2, K0 star with a metallicity of [Fe/H] = +0.28 $\pm$ 0.13. It appears to be old, with an expanded radius for a low-mass star ($R$ = 1.11 $\pm$ 0.05 R$_{\odot}$; $M$ = 0.95 $\pm$ 0.05  M$_{\odot}$) leading to an age estimate of 13 $\pm$ 2 Gy.    This star may have an inflated
radius owing to magnetic activity. This phenomenon can be accounted for to some
extent by using models with a lower mixing-length parameter ($\alpha_{\rm MLT}$). If we assume $\alpha_{\rm MLT} = 1.50$ for this star instead of
standard value of $\alpha_{\rm MLT} = 1.78$ from a solar calibration, we
obtain a best-fit age of 9.2\,Gyr.

The planet is massive, at 5.2 $\pm$ 0.2 M$_{\rm Jup}$, and is in a relatively long and eccentric orbit ($P$ = 9.6 d, $e$ = 0.43; Fig.~9; Table~7).   The circularisation timescale for such a planet (e.g. eqn 3 of \citet{2006ApJ...649.1004A}, and assuming $Q_{P}$ $\sim$ 10$^{5}$) would be of order 30 Gyr, and so the eccentricity is compatible with the old age of the star.    Given the transit, the probability that there is also an occultation of the planet (using eqn 2 of \citealt{2009PASP..121.1096K}) is greater than 0.46. 

Comparable high-mass, long-period hot Jupiters in eccentric orbits include WASP-8b (2.2 M$_{\rm Jup}$, 8.2 d, $e$ = 0.31; \citealt{2010A&A...517L...1Q}) and Kepler-75 (9.9 M$_{\rm Jup}$, 8.9 d,  $e$ = 0.57; \citealt{2013A&A...554A.114H}).

\begin{figure}
\hspace*{2mm}\includegraphics[width=8.5cm]{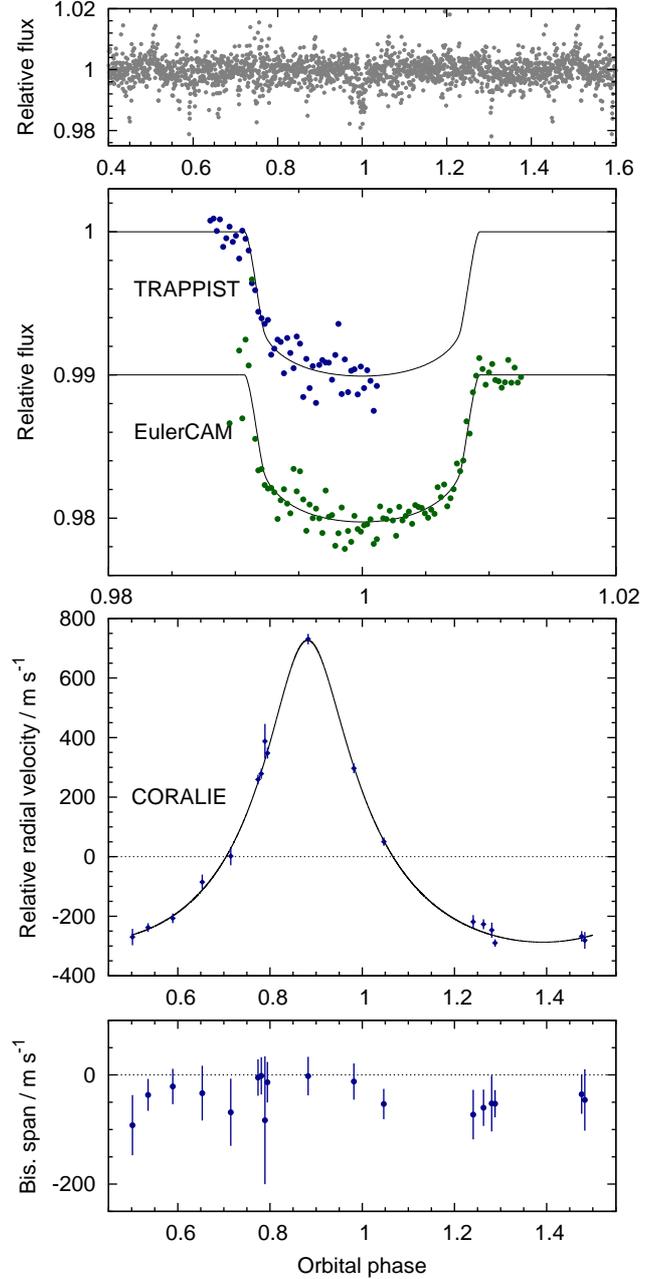}\\ [-2mm]
\caption{WASP-162b discovery data, as for Figs.~2 \&\ 3.}
\end{figure}

\begin{table}
\caption{System parameters for WASP-168.}  
\begin{tabular}{lc}
\multicolumn{2}{l}{1SWASP\,062658.70--464917.1}\\
\multicolumn{2}{l}{2MASS\,06265871-4649171}\\
\multicolumn{2}{l}{GAIA RA\,=\,06$^{\rm h}$26$^{\rm m}$58.71$^{\rm s}$, 
Dec\,=\,--46$^{\circ}$49$^{'}$17.2$^{''}$ (J2000)}\\
\multicolumn{2}{l}{$V$ mag = 12.1}  \\ 
\multicolumn{2}{l}{Rotational modulation\ \ \ $<$ 3 mmag  (95\%)}\\
\multicolumn{2}{l}{UCAC5 pm (RA) 0.0\,$\pm$\,1.1 (Dec) 20.1\,$\pm$\,1.1 mas/yr}\\
\multicolumn{2}{l}{GAIA parallax: 3.13  $\pm$ 0.25 mas}\\
\hline
\multicolumn{2}{l}{Stellar parameters from spectroscopic analysis.\rule[-1.5mm]{0mm}{2mm}} \\ \hline 
Spectral type & F9V \\
$T_{\rm eff}$ (K)  & 6000  $\pm$ 100  \\
$\log g$      & 4.0 $\pm$ 0.1    \\
$v\,\sin i$ (km\,s$^{-1}$)     &    0.3 $\pm$ 0.1     \\
{[Fe/H]}   &   $-$0.01 $\pm$ 0.09     \\
log A(Li)  &    2.93 $\pm$ 0.12      \\
Age (Lithium) [Gy]  &     $<$ 1       \\ \hline 
\multicolumn{2}{l}{Parameters from MCMC analysis.\rule[-1.5mm]{0mm}{3mm}} \\
\hline 
$P$ (d) & 4.153658  $\pm$ 0.000003 \\
$T_{\rm c}$ (HJD)\,(UTC) & 245\,7424.5278 $\pm$ 0.0004 \\
$T_{\rm 14}$ (d) & 0.0797 $\pm$ 0.0017 \\
$T_{\rm 12}=T_{\rm 34}$ (d) & undefined \\
$\Delta F=R_{\rm P}^{2}$/R$_{*}^{2}$ & 0.0119 $^{+0.0015}_{-0.0007}$\\
$b$ & 0.97 $^{+0.06}_{-0.04}$ \\
$i$ ($^\circ$)  & 84.4 $\pm$ 0.6 \\
$K_{\rm 1}$ (km s$^{-1}$) & 0.050 $\pm$ 0.004 \\
$\gamma$ (km s$^{-1}$)  &50.460  $\pm$ 0.003 \\
$e$ & 0 (adopted) ($<$\,0.09 at 2$\sigma$) \\ 
$a/R_{\rm *}$  & $9.59^{+0.65}_{-0.39}$\\
$M_{\rm *}$ (M$_{\rm \odot}$) & 1.08 $\pm$ 0.05 \\
$R_{\rm *}$ (R$_{\rm \odot}$) & 1.12 $\pm$ 0.06 \\
$\log g_{*}$ (cgs) & 4.37 $\pm$ 0.05  \\
$\rho_{\rm *}$ ($\rho_{\rm \odot}$) & 0.77 $\pm$ 0.14\\
$T_{\rm eff}$ (K) & 6000 $\pm$ 100 \\
$M_{\rm P}$ (M$_{\rm Jup}$) & 0.42 $\pm$ 0.04 \\
$R_{\rm P}$ (R$_{\rm Jup}$) & 1.5 $^{+0.5}_{-0.3}$\\ 
$\log g_{\rm P}$ (cgs) & 2.6 $\pm$ 0.3 \\
$\rho_{\rm P}$ ($\rho_{\rm J}$) & 0.12 $^{+0.10}_{-0.07}$  \\
$a$ (AU)  & 0.0519   $\pm$ 0.0008 \\
$T_{\rm P, A=0}$ (K) & 1340  $\pm$ 40 \\ [0.5mm] \hline 
\multicolumn{2}{l}{Errors are 1$\sigma$; Limb-darkening coefficients were:}\\
\multicolumn{2}{l}{{\small $r$ band: a1 = 0.547, a2 = 0.084, a3 = 0.308, 
a4 = --0.215}}\\ 
\multicolumn{2}{l}{{\small $z$ band: a1 =   0.633 , a2 = --0.263 , a3 = 0.523, a4 = --0.280}}\\ 
\multicolumn{2}{l}{{\small $V$ band: a1 =   0.462 , a2 = 0.310 , a3 = 0.181, a4 = --0.170}}\\ 
\multicolumn{2}{l}{{\small $I$ band: a1 =   0.627 , a2 = --0.208 , a3 = 0.518, a4 = --0.286}}\\ \hline
\end{tabular} 
\end{table}

\section{WASP-168}
WASP-168  (Figs.~10 \&\ 11; Table~8) is a $V$ = 12.1, F9 star with a metallicity of [Fe/H] = $-$0.01 $\pm$ 0.09. The evolutionary tracks indicate an age of 4 $\pm$ 2 Gy.   The lithium abundance of log A(Li) 2.93 $\pm$ 0.12 indicates an age of $<$ 1 Gy, though we consider this less reliable. 

The planet has a 4.15-d orbit.  Of the three follow-up transit lightcurves, note that the EulerCAM transit was observed simultaneously by TRAPPIST. These all show a grazing transit with a high impact factor ($b$ = 0.97 $\pm$ 0.05).   As with WASP-145Ab, this means that the 2$^{\rm nd}$ and 3$^{\rm rd}$ contact are not discernable and thus that the planetary radius is not well constrained. Nevertheless, the estimates give a bloated planet with a mass of 0.42 $\pm$ 0.04 M$_{\rm Jup}$ and a radius of 1.5 $^{+0.5}_{-0.3}$ R$_{\rm Jup}$. As noted for WASP-159b, bloated hot Jupiters with similar parameters are commonly found.

\begin{figure}
\hspace*{2mm}\includegraphics[width=8.5cm]{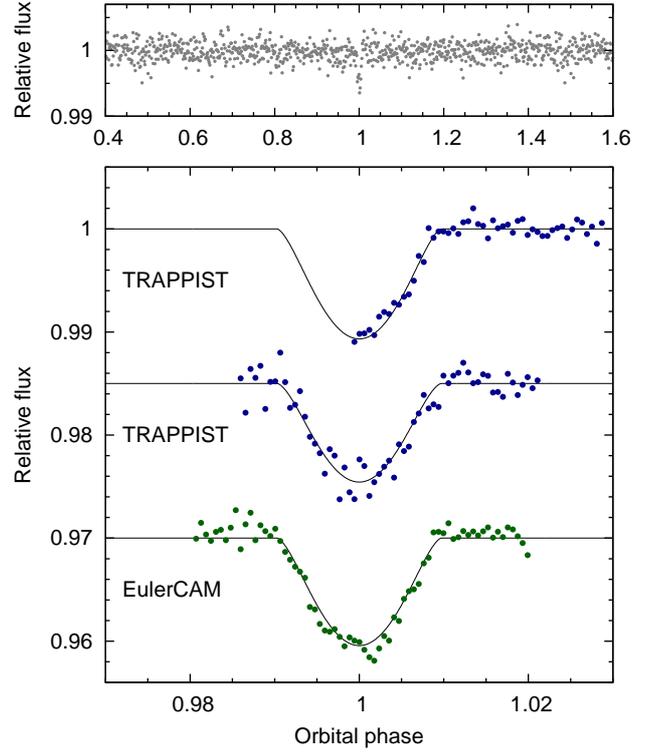}\\ [-2mm]
\caption{WASP-168b discovery photometry, as for Fig.~2.}
\end{figure}

\begin{figure}
\hspace*{2mm}\includegraphics[width=8.5cm]{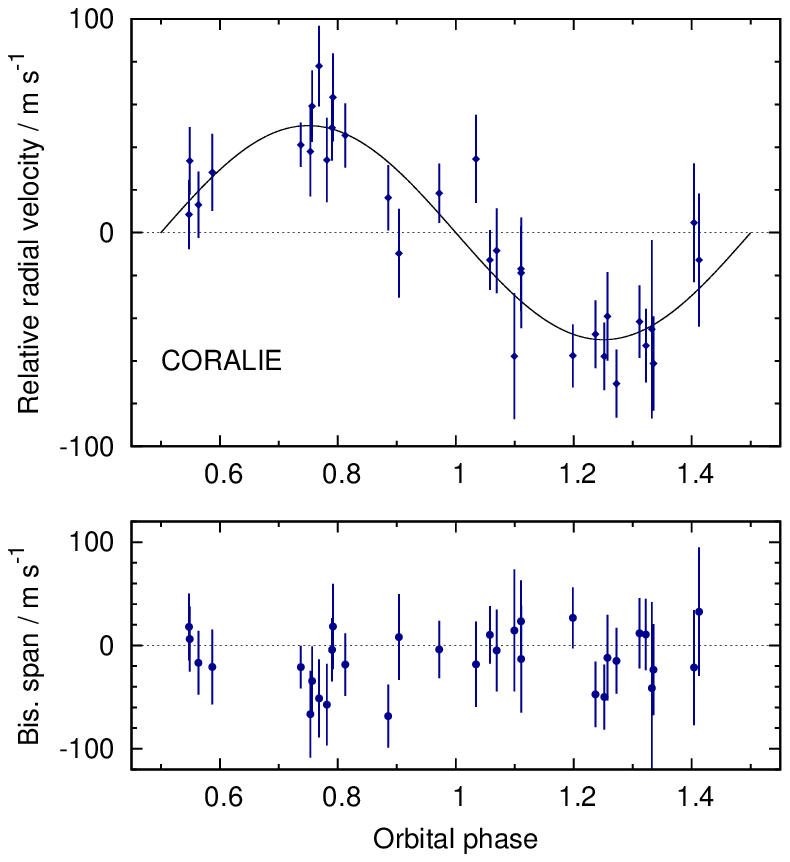}\\ [-2mm]
\caption{WASP-168b radial velocities and bisector spans, as for Fig.~3.}
\end{figure}

\clearpage

\begin{table}
\caption{System parameters for WASP-172.}  
\begin{tabular}{lc}
\multicolumn{2}{l}{1SWASP\,131744.13--471415.3}\\
\multicolumn{2}{l}{2MASS\,13174412--4714152}\\
\multicolumn{2}{l}{GAIA RA\,=\,13$^{\rm h}$17$^{\rm m}$44.12$^{\rm s}$, 
Dec\,=\,--47$^{\circ}$14$^{'}$15.3$^{''}$ (J2000)}\\
\multicolumn{2}{l}{$V$ mag = 11.0}  \\ 
\multicolumn{2}{l}{Rotational modulation: $<$ 1 mmag (95\%)}\\
\multicolumn{2}{l}{UCAC5 pm (RA) --16.2\,$\pm$\,0.8 (Dec) 0.7\,$\pm$\,0.8 mas/yr}\\ 
\multicolumn{2}{l}{GAIA parallax: 1.78  $\pm$ 0.26 mas}\\
\hline
\multicolumn{2}{l}{Stellar parameters from spectroscopic analysis.\rule[-1.5mm]{0mm}{2mm}} \\ \hline 
Spectral type & F1V \\
$T_{\rm eff}$ (K)  & 6900  $\pm$ 150  \\
$\log g$      & 4.1 $\pm$ 0.2    \\
$v\,\sin i$ (km\,s$^{-1}$)     &    13.7 $\pm$ 1.0     \\
{[Fe/H]}   &   $-$0.10 $\pm$ 0.08     \\
log A(Li)  &    $<$ 1.2      \\
Age (Lithium) [Gy]  &  (too hot for estimate)         \\ \hline 
\multicolumn{2}{l}{Parameters from MCMC analysis.\rule[-1.5mm]{0mm}{3mm}} \\
\hline 
$P$ (d) &  5.477433 $\pm$ 0.000007 \\
$T_{\rm c}$ (HJD)\,(UTC) & 245\,7032.2617 $\pm$ 0.0005 \\
$T_{\rm 14}$ (d) & 0.2206 $\pm$ 0.0020 \\
$T_{\rm 12}=T_{\rm 34}$ (d) & 0.021  $\pm$ 0.002\\
$\Delta F=R_{\rm P}^{2}$/R$_{*}^{2}$ & 0.0072 $\pm$ 0.0002\\
$b$ & 0.45 $\pm$ 0.12 \\
$i$ ($^\circ$)  & 86.7 $\pm$ 1.1 \\
$K_{\rm 1}$ (km s$^{-1}$) & 0.042 $\pm$ 0.009 \\
$\gamma$ (km s$^{-1}$)  & --20.283 $\pm$ 0.006 \\
$e$ & 0 (adopted) ($<$\,0.28 at 2$\sigma$) \\ 
$a/R_{\rm *}$  & 8.0 $\pm$ 0.5\\
$M_{\rm *}$ (M$_{\rm \odot}$) & 1.49 $\pm$ 0.07 \\
$R_{\rm *}$ (R$_{\rm \odot}$) & 1.91 $\pm$ 0.10 \\
$\log g_{*}$ (cgs) & 4.05 $\pm$ 0.05  \\
$\rho_{\rm *}$ ($\rho_{\rm \odot}$) & 0.21 $\pm$ 0.04\\
$T_{\rm eff}$ (K) & 6900 $\pm$ 140 \\
$M_{\rm P}$ (M$_{\rm Jup}$) & 0.47 $\pm$ 0.10 \\
$R_{\rm P}$ (R$_{\rm Jup}$) & 1.57 $\pm$ 0.10 \\ 
$\log g_{\rm P}$ (cgs) & 2.64 $\pm$ 0.11 \\
$\rho_{\rm P}$ ($\rho_{\rm J}$) & 0.12 $\pm$ 0.04 \\
$a$ (AU)  & 0.0694   $\pm$ 0.0011 \\
$T_{\rm P, A=0}$ (K) & 1740  $\pm$ 60 \\ [0.5mm] \hline 
\multicolumn{2}{l}{Errors are 1$\sigma$; Limb-darkening coefficients were:}\\
\multicolumn{2}{l}{{\small $r$ band: a1 =  0.422, a2 = 0.603, a3 = --0.477, 
a4 = 0.129}}\\ 
\multicolumn{2}{l}{{\small $z$ band: a1 =   0.513 , a2 = 0.159, , a3 = --0.089, a4 = --0.018}}\\ \hline
\end{tabular} 
\end{table}

\section{WASP-172}
WASP-172 is a fairly hot, F1 star with $V$ = 11.0 and a metallicity of [Fe/H] =   $-$0.10 $\pm$ 0.08. The age estimate is 1.8 $\pm$ 0.3 Gy.   WASP-172b is a moderately bloated planet (0.5 M$_{\rm Jup}$; 1.6 R$_{\rm Jup}$) in a 5.48-d orbit  (Figs.~12 \&\ 13; Table~9).   

The large stellar radius (1.9 R$_{\odot}$) leads to a small transit depth of 0.7\%\, and thus a ground-based survey would struggle to have detected the planet if it were not bloated.   Thus, while highly irradiated hot Jupiters around hot stars are often found to be bloated one has to be careful with observational biases in constructing such samples.  

\begin{figure}
\hspace*{2mm}\includegraphics[width=8.5cm]{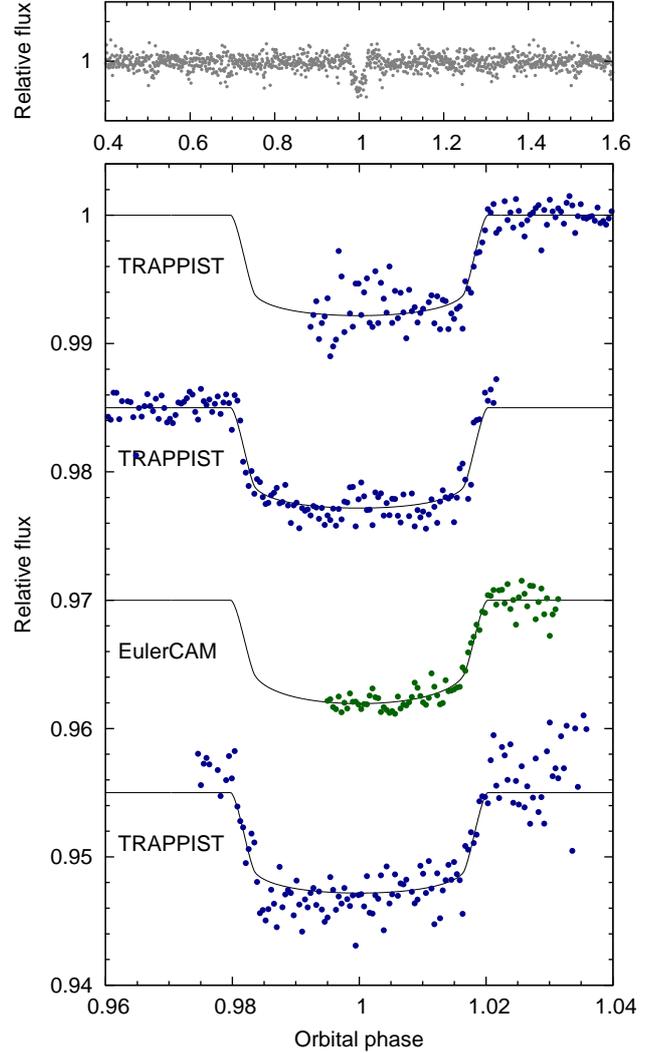}\\ [-2mm]
\caption{WASP-172b discovery photometry, as for Fig.~2.}
\end{figure}

\begin{figure}
\hspace*{2mm}\includegraphics[width=8.5cm]{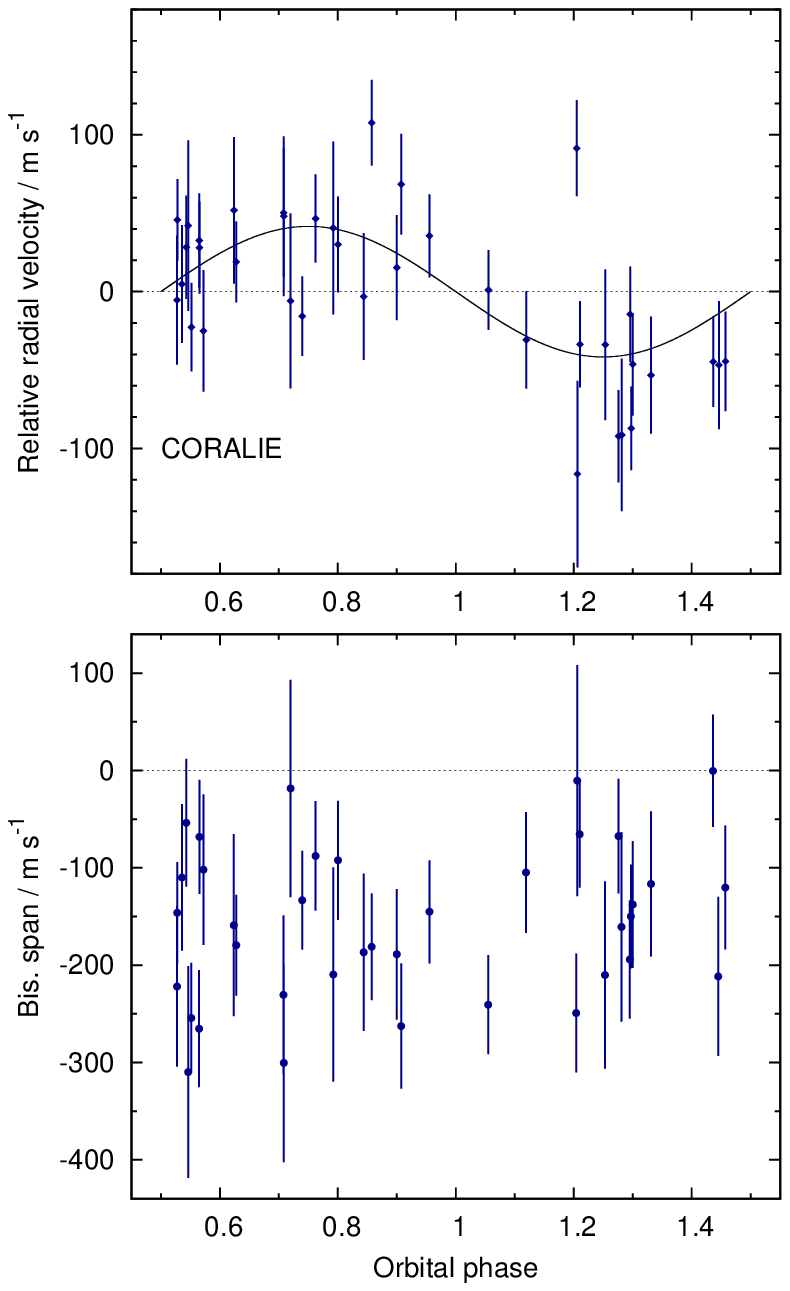}\\ [-2mm]
\caption{WASP-172b radial velocities and bisector spans, as for Fig.~3.}
\end{figure}

\begin{table}
\caption{System parameters for WASP-173.}  
\begin{tabular}{lc}
\multicolumn{2}{l}{1SWASP\,233640.32--343640.4}\\
\multicolumn{2}{l}{2MASS\,23364036--3436404}\\
\multicolumn{2}{l}{WDS23366$-$3437}\\
\multicolumn{2}{l}{GAIA RA\,=\,23$^{\rm h}$36$^{\rm m}$40.38$^{\rm s}$, 
Dec\,=\,--34$^{\circ}$36$^{'}$40.6$^{''}$ (J2000)}\\
\multicolumn{2}{l}{$V$ mag = 11.3}  \\ 
\multicolumn{2}{l}{Rotational modulation: $P$ = 7.9 $\pm$ 0.1; 6--15 mmag amplitude}\\
\multicolumn{2}{l}{UCAC5 pm (RA) 88.1\,$\pm$\,0.8 (Dec) --8.3\,$\pm$\,0.8 mas/yr}\\ 
\multicolumn{2}{l}{GAIA parallax: 4.34  $\pm$ 0.61 mas}\\
\hline
\multicolumn{2}{l}{Stellar parameters from spectroscopic analysis.\rule[-1.5mm]{0mm}{2mm}} \\ \hline 
Spectral type & G3 \\
$T_{\rm eff}$ (K)  & 5700  $\pm$ 150  \\
$\log g$      & 4.5 $\pm$ 0.2    \\
$v\,\sin i$ (km\,s$^{-1}$)     &    6.1 $\pm$ 0.3     \\
{[Fe/H]}   &  +0.16 $\pm$ 0.14     \\
log A(Li)  &    $<$ 0.8      \\
Age (Lithium) [Gy]  &  $>$ 5          \\  \hline 
\multicolumn{2}{l}{Parameters from MCMC analysis.\rule[-1.5mm]{0mm}{3mm}} \\
\hline 
$P$ (d) &  1.38665318 $\pm$ 0.00000027 \\
$T_{\rm c}$ (HJD)\,(UTC) & 245\,7288.8585 $\pm$ 0.0002 \\
$T_{\rm 14}$ (d) & 0.0957 $\pm$ 0.0007 \\
$T_{\rm 12}=T_{\rm 34}$ (d) & 0.0117  $\pm$ 0.0009\\
$\Delta F=R_{\rm P}^{2}$/R$_{*}^{2}$ & 0.0123 $\pm$ 0.0002\\
$b$ & 0.40 $\pm$ 0.08 \\
$i$ ($^\circ$)  & 85.2 $\pm$ 1.1 \\
$K_{\rm 1}$ (km s$^{-1}$) & 0.645 $\pm$ 0.007 \\
$\gamma$ (km s$^{-1}$)  & --7.858$\pm$ 0.004 \\
$e$ & 0 (adopted) ($<$\,0.032 at 2$\sigma$) \\
$a/R_{\rm *}$  & 4.78 $\pm$ 0.17\\
$M_{\rm *}$ (M$_{\rm \odot}$) & 1.05 $\pm$ 0.08 \\
$R_{\rm *}$ (R$_{\rm \odot}$) & 1.11 $\pm$ 0.05 \\
$\log g_{*}$ (cgs) & 4.37 $\pm$ 0.03  \\
$\rho_{\rm *}$ ($\rho_{\rm \odot}$) & 0.76 $\pm$ 0.08\\
$T_{\rm eff}$ (K) & 5800 $\pm$ 140 \\
$M_{\rm P}$ (M$_{\rm Jup}$) & 3.69 $\pm$ 0.18 \\
$R_{\rm P}$ (R$_{\rm Jup}$) & 1.20 $\pm$ 0.06 \\ 
$\log g_{\rm P}$ (cgs) & 3.77 $\pm$ 0.04 \\
$\rho_{\rm P}$ ($\rho_{\rm J}$) & 2.14 $\pm$ 0.28 \\
$a$ (AU)  & 0.0248   $\pm$ 0.0006 \\
$T_{\rm P, A=0}$ (K) & 1880  $\pm$ 55 \\ [0.5mm] \hline 
\multicolumn{2}{l}{Errors are 1$\sigma$; Limb-darkening coefficients were:}\\
\multicolumn{2}{l}{{\small $r$ band: a1 = 0.673, a2 = --0.360, a3 = 0.893, 
a4 = --0.448}}\\ 
\multicolumn{2}{l}{{\small $z$ band: a1 =   0.754, a2 = --0.625 , a3 = 0.957, a4 = --0.442}}\\ 
\multicolumn{2}{l}{{\small $V$ band: a1 =   0.602, a2 = --0.228 , a3 = 0.902, a4 = --0.459}}\\ \hline
\end{tabular} 
\end{table}

\section{WASP-173}
WASP-173A  (Figs.~14 \&\ 15; Table~10) is the brighter component of a known double-star system catalogued as WDS23366$-$3437 \citep{2001AJ....122.3466M}. The two components have $V$ = 11.3 and 12.1 and are separated by 6 arcsecs. Given the GAIA parallax of 4.34  $\pm$ 0.61 mas this amounts to a separation 1400 $\pm$ 200 AU.   Our spectral analysis suggests a type of G3 and a metallicity of +0.16 $\pm$ 0.14 for WASP-173A (we do not have spectra of the companion).     The age estimate is 7 $\pm$ 3 Gy.  

The WASP data show a rotational modulation at a period of 7.9 $\pm$ 0.1 d (Fig.~16).  It is seen independently in three seasons (2006, 2007, 2011), with false-alarm probabilities of  $<$0.1\%\ in two of them.  The amplitude ranges from 6 to 15 mmag.     Given the 14-arcsec pixels of WASP data we cannot tell which of the two stars is producing the modulation, but the amplitude suggests that it is more likely to be the brighter component, WASP-173A (note that the quoted amplitude has not been corrected for dilution).   The follow-up transit lightcurves with EulerCAM and TRAPPIST used a smaller aperture containing only WASP-173A.  

WASP-173Ab is a massive planet (3.7 M$_{\rm Jup}$) in a close-in, circular  orbit with a period of  1.39 d.  The radius is 1.2 R$_{\rm Jup}$.

Given the stellar radius of 1.11 $\pm$ 0.05 R$_{\rm \odot}$ the rotation period implies a surface velocity of 7.15 $\pm$ 0.35  km s$^{-1}$. This compares with the measured \vsini\ of  6.1 $\pm$ 0.3 km s$^{-1}$. This suggests that the spin axis might not be fully perpendicular to us, and thus that the planet might be in a moderately misaligned orbit.  This system may be a worthwhile target for an observation of the Rossiter--McLaughlin effect.

\begin{figure}
\hspace*{2mm}\includegraphics[width=8.5cm]{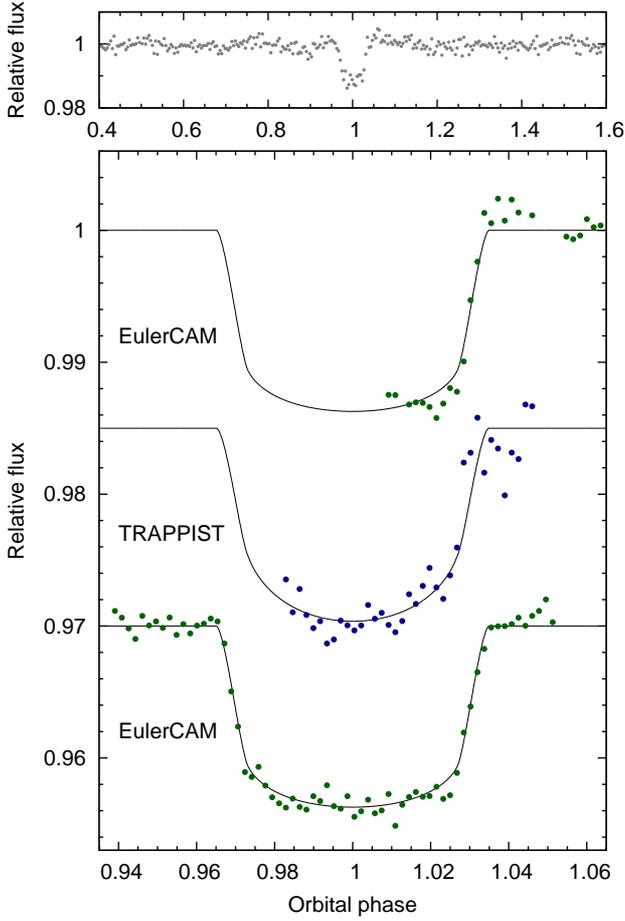}\\ [-2mm]
\caption{WASP-173b discovery photometry, as for Fig.~2.}
\end{figure}

\begin{figure}
\hspace*{2mm}\includegraphics[width=8.5cm]{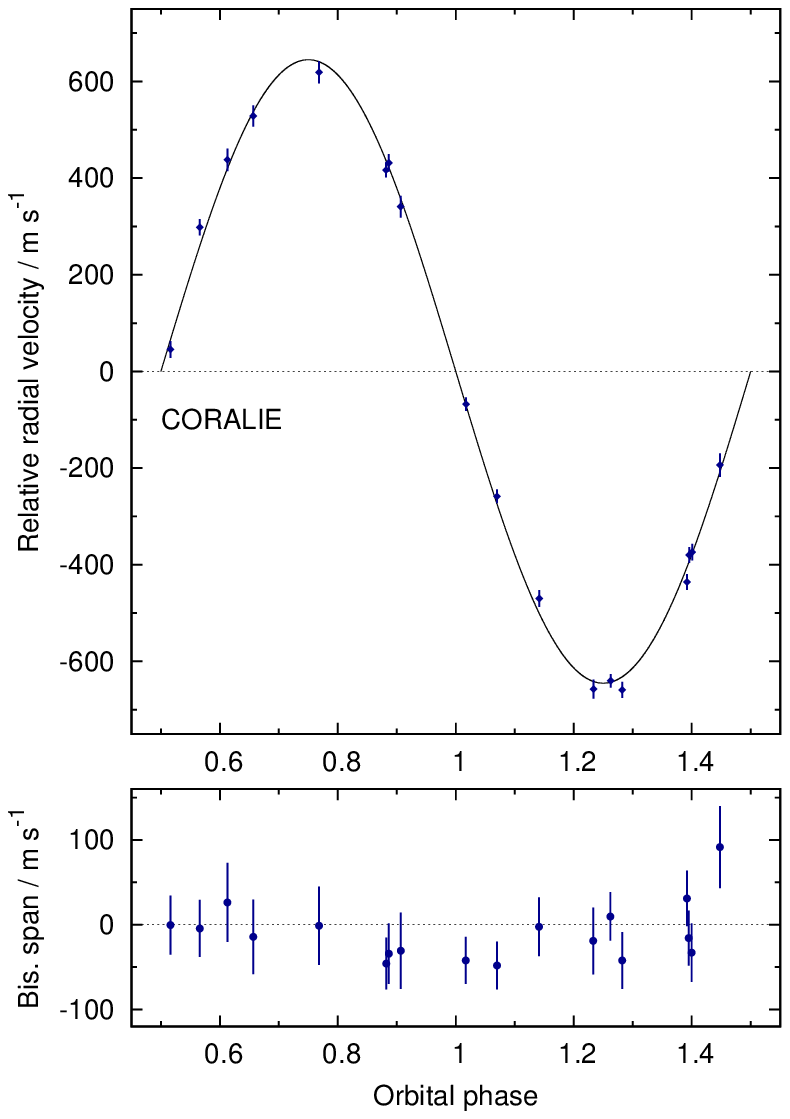}\\ [-2mm]
\caption{WASP-173b radial velocities and bisector spans, as for Fig.~3.}
\end{figure}

\begin{figure}
\hspace*{0mm}\includegraphics[width=9cm]{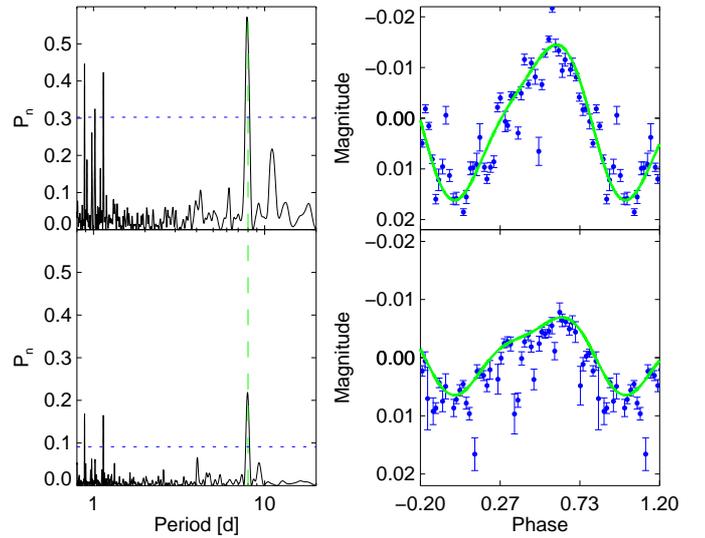}\\ [-2mm]
\caption{WASP-173 rotational modulation. The left-hand panels show periodograms for independent WASP data sets  from 2007 and 2011. The blue dotted line is a false-alarm probability of 0.001.  The right-hand panels show the data for each season folded on the 7.9-d period; the green line is a harmonic-series fit to the data}
\end{figure}

\section{Conclusions}
With regards to atmospheric characterisation the most promising of the new targets presented here is WASP-172b.  This is a bloated, low-mass hot Jupiter (0.47 M$_{\rm Jup}$; 1.57  R$_{\rm Jup}$) transiting a relatively bright star at $V$ = 11.0.   The atmosphere is also predicted to be hot (1740 $\pm$ 60 K) owing to the hot, F1 host star.    Other bloated planets in this batch are WASP-159b, but that has a fainter host with $V$ = 12.9, and WASP-168b, which has both a fainter host with $V$ = 12.1 and a grazing transit making it harder to parametrise.  

WASP-144b is notable for being un-bloated ($R$ = 0.85 $\pm$ 0.05 R$_{\rm Jup}$). WASP-145b may also be unusually small for a hot Jupiter, but this is uncertain owing to the transit being grazing.  

Three of the planets reported here are relatively massive, WASP-158b at 2.8 M$_{\rm Jup}$, WASP-173Ab at 3.7 M$_{\rm Jup}$, and WASP-162b at 5.2 M$_{\rm Jup}$. Such planets are relatively less common.  Before this paper, only 21 transiting hot Jupiters were known with masses in the range 2.5--6 M$_{\rm Jup}$ compared to 205 in the range 0.5--2.5  M$_{\rm Jup}$.   Of the three new massive planets, only WASP-162b has a significant eccentricity ($e$ = 0.43). 

Two of the planets reported here (WASP-145Ab and WASP-173Ab) are in double-star systems.   

Lastly, our results show that the combination of WASP-South, CORALIE and TRAPPIST continues to be a productive team for discovering hot Jupiters transiting stars of $V < 13$. 

\section*{Acknowledgements}
WASP-South is hosted by the South African Astronomical Observatory and
we are grateful for their ongoing support and assistance. Funding for
WASP comes from consortium universities and from the UK's Science and
Technology Facilities Council. The Euler Swiss telescope is supported
by the Swiss National Science Foundation. TRAPPIST is funded by the
Belgian Fund for Scientific Research (Fond National de la Recherche
Scientifique, FNRS) under the grant FRFC 2.5.594.09.F, with the
participation of the Swiss National Science Fundation (SNF). L.D. 
acknowledges support from a Gruber Foundation Fellowship.

\bibliographystyle{mnras}
\bibliography{biblio}




\begin{table}
\renewcommand\thetable{A1}
\caption{Radial velocities.\protect\rule[-1.5mm]{0mm}{2mm}} 
\begin{tabular}{cccr} 
\hline 
BJD\,--\,2400\,000 & RV & $\sigma_{\rm RV}$ & Bisector \\
(UTC)  & (km s$^{-1}$) & (km s$^{-1}$) & (km s$^{-1}$)\\ [0.5mm] \hline
\multicolumn{4}{l}{{\bf WASP-144:}}\\  
56837.74442 & 16.1865 & 0.0468 & $-$0.0628 \\
56845.73476 & 16.0438 & 0.0282 & $-$0.0215 \\
56926.63966 & 16.1629 & 0.0189 & 0.0062	 \\
56927.64776 & 16.0215 & 0.0220 & 0.0326	 \\
56928.60109 & 16.1807 & 0.0241 & 0.1021	 \\
56930.71731 & 16.1352 & 0.0244 & 0.1589	 \\
56931.73201 & 16.0526 & 0.0324 & 0.0036	 \\
56942.60950 & 16.2039 & 0.0271 & 0.0623	 \\
56955.60807 & 16.1140 & 0.0251 & $-$0.0947 \\
56958.62456 & 16.1967 & 0.0474 & $-$0.0913 \\ \cline{1-1} 
56983.52692 & 16.1859 & 0.0398 & $-$0.0585 \\
57140.88455 & 16.2049 & 0.0434 & 0.0129	 \\
57325.59467 & 16.1622 & 0.0510 & $-$0.0276 \\
57576.93050 & 16.0157 & 0.0410 & $-$0.0893 \\
57584.76134 & 16.1852 & 0.0308 & $-$0.0615 \\
57664.69172 & 16.1616 & 0.0256 & 0.0079	 \\
57712.56278 & 16.2271 & 0.0761 & $-$0.0750 \\ [0.5mm] 
\hline
\multicolumn{4}{l}{Bisector errors are twice RV errors} 
\end{tabular} 
\end{table}

\newpage 

\vspace*{-2cm}

\begin{tabular}{cccr} 
\hline 
BJD\,--\,2400\,000 & RV & $\sigma_{\rm RV}$ & Bisector \\
(UTC)  & (km s$^{-1}$) & (km s$^{-1}$) & (km s$^{-1}$)\\ [0.5mm] \hline
\multicolumn{4}{l}{{\bf WASP-145:}}\\  
56811.88123 & 3.5320 & 0.0120 & 0.0275 \\ 
56890.59823 & 3.1457 & 0.0173 & 0.0372 \\ 
56903.76315 & 3.4954 & 0.0218 & 0.0133 \\ 
56916.59725 & 3.4016 & 0.0184 & 0.0666 \\ 
56927.67549 & 3.1723 & 0.0117 & 0.0048 \\ 
56928.65445 & 3.5252 & 0.0117 & 0.0555 \\ 
56942.63948 & 3.4927 & 0.0127 & 0.0234 \\ 
56950.50317 & 3.2476 & 0.0128 & 0.0179 \\ 
56956.50488 & 3.3369 & 0.0182 & 0.0057 \\ \cline{1-1} 
56996.55409 & 3.1886 & 0.0164 & -0.057 \\ 
57140.91392 & 3.5285 & 0.0176 & 0.0123 \\ 
57226.68125 & 3.1481 & 0.0368 & 0.0345 \\ 
57295.70398 & 3.1732 & 0.0164 & 0.0375 \\ 
57325.62279 & 3.2030 & 0.0242 & 0.0155 \\ 
57493.89365 & 3.1607 & 0.0118 & -0.042 \\ 
57569.71114 & 3.2203 & 0.0146 & 0.0364 \\ 
57584.85526 & 3.5170 & 0.0134 & 0.0052 \\ 
57600.73014 & 3.5075 & 0.0253 & -0.064 \\ 
57629.57199 & 3.3861 & 0.0183 & -0.028 \\ [0.5mm]
\multicolumn{4}{l}{{\bf WASP-158:}}\\ 
 56876.83406 & 24.4864 & 0.0402 & $-$0.0026 \\
57006.53670 & 23.9180 & 0.0557 & 0.1307	 \\
57007.59318 & 24.3047 & 0.0550 & $-$0.1263 \\
57008.60038 & 24.4297 & 0.0516 & $-$0.1734 \\
57010.57920 & 23.9154 & 0.0447 & 0.0440	 \\
57011.56783 & 24.4198 & 0.0492 & $-$0.0777 \\
57014.58938 & 24.1044 & 0.0440 & 0.0382	 \\
57203.89522 & 23.8789 & 0.0629 & $-$0.0082 \\
57205.82496 & 24.6116 & 0.0874 & 0.1649	 \\
57224.78600 & 24.2981 & 0.1224 & 0.1592	 \\
57255.72861 & 24.0305 & 0.0740 & $-$0.0932 \\
57302.75125 & 23.9124 & 0.1391 & 0.0988	 \\
57327.70752 & 23.9901 & 0.1039 & 0.0236	 \\
57370.56015 & 24.4935 & 0.0647 & 0.0285	 \\
57559.88888 & 24.3833 & 0.0641 & 0.2210	 \\
57587.77799 & 23.9875 & 0.0552 & $-$0.0663 \\
57598.84959 & 23.9258 & 0.0478 & $-$0.0213 \\
57613.72822 & 23.8843 & 0.0362 & 0.0952	 \\
57655.75797 & 24.4331 & 0.0518 & 0.0073	 \\
57692.71279 & 24.3341 & 0.0619 & $-$0.1398 \\ [0.5mm] 
\hline
\multicolumn{4}{l}{Bisector errors are twice RV errors} 
\end{tabular} 



\begin{tabular}{cccr} 
\hline 
BJD\,--\,2400\,000 & RV & $\sigma_{\rm RV}$ & Bisector \\
(UTC)  & (km s$^{-1}$) & (km s$^{-1}$) & (km s$^{-1}$)\\ [0.5mm] \hline
\multicolumn{4}{l}{\bf WASP-159:}\\  
56979.72275 & 35.1927 & 0.0392 & 0.0045 \\ 
57004.70593 & 35.1071 & 0.0370 & $-$0.0080 \\ 
57022.73682 & 35.1871 & 0.0320 & 0.0114 \\ 
57033.65716 & 35.2266 & 0.0190 & 0.0563 \\ 
57035.55510 & 35.0952 & 0.0291 & $-$0.0921 \\ 
57036.69928 & 35.2257 & 0.0272 & 0.0252 \\ 
57061.57339 & 35.1507 & 0.0567 & 0.2588 \\ 
57086.56873 & 35.1548 & 0.0242 & 0.0820 \\ 
57089.53336 & 35.1237 & 0.0315 & 0.1049 \\ 
57261.90726 & 35.0649 & 0.0432 & 0.0070 \\ 
57286.75364 & 35.1981 & 0.0679 & $-$0.1212 \\ 
57331.79414 & 35.1973 & 0.0594 & 0.1496 \\ 
57333.76731 & 35.1721 & 0.0343 & 0.0166 \\ 
57338.78070 & 35.1030 & 0.0486 & 0.1369 \\ 
57340.81000 & 35.1969 & 0.0391 & 0.0116 \\ 
57367.67376 & 35.1806 & 0.0369 & 0.0566 \\ 
57388.60652 & 35.1290 & 0.0325 & 0.0089 \\ 
57390.66011 & 35.2368 & 0.0256 & $-$0.0008 \\ 
57398.60368 & 35.1855 & 0.0282 & 0.0600 \\ 
57399.64861 & 35.1485 & 0.0356 & $-$0.0326 \\ 
57416.62863 & 35.1671 & 0.0354 & $-$0.0714 \\ 
57429.55929 & 35.2009 & 0.0189 & 0.0197 \\ 
57689.81060 & 35.2180 & 0.0221 & 0.0669 \\ 
57715.79980 & 35.1126 & 0.0517 & 0.2065 \\ 
57726.65988 & 35.0920 & 0.0280 & $-$0.0658 \\ 
57751.67104 & 35.1994 & 0.0339 & $-$0.0262 \\ 
57770.62964 & 35.2220 & 0.0247 & $-$0.0469 \\ 
57806.59954 & 35.1281 & 0.0394 & 0.0558 \\ 
57823.55477 & 35.1264 & 0.0294 & $-$0.0393 \\ [0.5mm] 
\hline
\multicolumn{4}{l}{Bisector errors are twice RV errors} 
\end{tabular}

\begin{tabular}{cccr} 
\hline 
BJD\,--\,2400\,000 & RV & $\sigma_{\rm RV}$ & Bisector \\
(UTC)  & (km s$^{-1}$) & (km s$^{-1}$) & (km s$^{-1}$)\\ [0.5mm] \hline
\multicolumn{4}{l}{{\bf WASP-162:}}\\  
56770.55906 & 16.5348 & 0.0125 & $-$0.0527 \\
57188.52985 & 16.8266 & 0.0307 & $-$0.0684 \\ 
57208.49314 & 17.2122 & 0.0586 & $-$0.0831 \\ 
57366.83385 & 16.6059 & 0.0227 & $-$0.0727 \\ 
57376.84521 & 16.5777 & 0.0256 & $-$0.0523 \\ 
57378.78780 & 16.5437 & 0.0281 & $-$0.0457 \\ 
57379.81346 & 16.6176 & 0.0162 & $-$0.0212 \\ 
57407.84558 & 16.5549 & 0.0275 & $-$0.0920 \\ 
57422.71696 & 16.8750 & 0.0139 & $-$0.0533 \\ 
57429.78462 & 17.1033 & 0.0171 & $-$0.0017 \\ 
57430.76149 & 17.5554 & 0.0176 & $-$0.0021 \\ 
57431.71535 & 17.1215 & 0.0166 & $-$0.0121 \\ 
57487.66160 & 17.1725 & 0.0186 & $-$0.0134 \\ 
57559.54335 & 16.5979 & 0.0167 & $-$0.0601 \\ 
57590.46580 & 16.5571 & 0.0178 & $-$0.0355 \\ 
57814.70374 & 17.0837 & 0.0167 & $-$0.0047 \\ 
57831.65846 & 16.5867 & 0.0146 & $-$0.0366 \\ 
57890.53621 & 16.7395 & 0.0251 & $-$0.0334 \\ [0.5mm]
\hline
\multicolumn{4}{l}{Bisector errors are twice RV errors} 
\end{tabular} 

\begin{tabular}{cccr} 
\hline 
BJD\,--\,2400\,000 & RV & $\sigma_{\rm RV}$ & Bisector \\
(UTC)  & (km s$^{-1}$) & (km s$^{-1}$) & (km s$^{-1}$)\\ [0.5mm] \hline
\multicolumn{4}{l}{{\bf WASP-168:}}\\  
56993.61587 & 50.4209 & 0.0207 & $-$0.0119 \\ 
57319.84988 & 50.5521 & 0.0650 & $-$0.0670 \\ 
57373.65989 & 50.4980 & 0.0201 & $-$0.0666 \\ 
57374.82620 & 50.4945 & 0.0206 & $-$0.0182 \\ 
57398.69788 & 50.4940 & 0.0197 & $-$0.0573 \\ 
57425.57334 & 50.4021 & 0.0158 & $-$0.0500 \\ 
57427.58986 & 50.5011 & 0.0103 & $-$0.0211 \\ 
57428.56442 & 50.4785 & 0.0138 & $-$0.0039 \\ 
57633.88933 & 50.4647 & 0.0279 & $-$0.0214 \\ 
57660.88579 & 50.4503 & 0.0208 &  0.0081	 \\ 
57661.74546 & 50.4431 & 0.0199 &  0.0232	 \\ 
57666.82069 & 50.4148 & 0.0417 & $-$0.0413 \\ 
57668.81582 & 50.5055 & 0.0151 & $-$0.0186 \\ 
57669.83349 & 50.4472 & 0.0140 &  0.0102	 \\ 
57671.87378 & 50.4937 & 0.0158 &  0.0062	 \\ 
57682.87937 & 50.4024 & 0.0148 &  0.0267	 \\ 
57694.80350 & 50.4516 & 0.0198 & $-$0.0049 \\ 
57699.80030 & 50.3893 & 0.0159 & $-$0.0149 \\ 
57715.69598 & 50.4021 & 0.0295 &  0.0145	 \\ 
57715.74476 & 50.4411 & 0.0259 & $-$0.0132 \\ 
57716.67587 & 50.3988 & 0.0221 & $-$0.0235 \\ 
57717.72294 & 50.4883 & 0.0180 & $-$0.0209 \\ 
57726.73304 & 50.5192 & 0.0168 & $-$0.0345 \\ 
57739.72958 & 50.4764 & 0.0152 & $-$0.0684 \\ 
57747.64158 & 50.5091 & 0.0153 & $-$0.0041 \\ 
57753.65099 & 50.4125 & 0.0159 & $-$0.0474 \\ 
57770.57567 & 50.4184 & 0.0170 &  0.0118	 \\ 
57800.63160 & 50.4685 & 0.0162 &  0.0179	 \\ 
57812.53164 & 50.4472 & 0.0311 &  0.0327	 \\ 
57824.61792 & 50.4071 & 0.0172 &  0.0106	 \\ 
57825.61993 & 50.4731 & 0.0155 & $-$0.0168 \\ 
57851.48939 & 50.5234 & 0.0207 &  0.0183	 \\ 
57855.54592 & 50.5380 & 0.0189 & $-$0.0512 \\ [0.5mm] 
\hline
\multicolumn{4}{l}{Bisector errors are twice RV errors} 
\end{tabular} 

\begin{tabular}{cccr} 
\hline 
BJD\,--\,2400\,000 & RV & $\sigma_{\rm RV}$ & Bisector \\
(UTC)  & (km s$^{-1}$) & (km s$^{-1}$) & (km s$^{-1}$)\\ [0.5mm] \hline
\multicolumn{4}{l}{{\bf WASP-172:}}\\  
56032.82587  & $-$20.2779  & 0.0376  & $-$0.1099  \\
56033.76992  & $-$20.2324  & 0.0408  & $-$0.2306  \\
56067.68648  & $-$20.2674  & 0.0335  & $-$0.1889  \\
56458.64357  & $-$20.3749  & 0.0295  & $-$0.0675  \\
56480.58391  & $-$20.3742  & 0.0486  & $-$0.1607  \\
56510.47908  & $-$20.2984  & 0.0254  & $-$0.1334  \\
56684.80361  & $-$20.2547  & 0.0293  & $-$0.0681  \\
56687.83916  & $-$20.3135  & 0.0311  & $-$0.1047  \\
56688.81248  & $-$20.3700  & 0.0266  & $-$0.1500  \\
56693.81486  & $-$20.3164  & 0.0275  & $-$0.0655  \\
56696.83656  & $-$20.2362  & 0.0282  & $-$0.0877  \\
56719.80357  & $-$20.2472  & 0.0265  & $-$0.1452  \\
56726.64829  & $-$20.1913  & 0.0306  & $-$0.2492  \\
56744.84931  & $-$20.2370  & 0.0260  & $-$0.1462  \\
56746.65439  & $-$20.1751  & 0.0274  & $-$0.1812  \\
56769.64903  & $-$20.2818  & 0.0255  & $-$0.2407  \\
56771.73503  & $-$20.3274  & 0.0289  & $-$0.0003  \\
56772.78389  & $-$20.2638  & 0.0259  & $-$0.1796  \\
56810.66120  & $-$20.2545  & 0.0329  & $-$0.0537  \\
56832.61861  & $-$20.3054  & 0.0283  & $-$0.2543  \\
56837.57872  & $-$20.3273  & 0.0318  & $-$0.1202  \\
56887.50487  & $-$20.3078  & 0.0387  & $-$0.1019  \\ \cline{1-1} 
57110.75884  & $-$20.3448  & 0.0373  & $-$0.1166  \\
57111.83467  & $-$20.2970  & 0.0411  & $-$0.2220  \\
57112.82654  & $-$20.2434  & 0.0510  & $-$0.3006  \\
57138.77562  & $-$20.3384  & 0.0409  & $-$0.2117  \\
57192.49655  & $-$20.3254  & 0.0481  & $-$0.2102  \\
57433.75786  & $-$20.3380  & 0.0326  & $-$0.1377  \\
57485.79471  & $-$20.2615  & 0.0305  & $-$0.0924  \\
57568.54383  & $-$20.2230  & 0.0321  & $-$0.2626  \\
57599.53201  & $-$20.2590  & 0.0301  & $-$0.2653  \\
57603.53378  & $-$20.3060  & 0.0304  & $-$0.1942  \\
57617.49253  & $-$20.2947  & 0.0404  & $-$0.1868  \\
57619.47611  & $-$20.4079  & 0.0594  & $-$0.0104  \\
57901.63900  & $-$20.2975  & 0.0558  & $-$0.0185  \\
57917.54541  & $-$20.2397  & 0.0468  & $-$0.1590  \\
57918.46887  & $-$20.2510  & 0.0551  & $-$0.2097  \\
57922.59777  & $-$20.2494  & 0.0544  & $-$0.3099  \\ [0.5mm] 
\hline
\multicolumn{4}{l}{Bisector errors are twice RV errors} 
\end{tabular}

\begin{tabular}{cccr} 
\hline 
BJD\,--\,2400\,000 & RV & $\sigma_{\rm RV}$ & Bisector \\
(UTC)  & (km s$^{-1}$) & (km s$^{-1}$) & (km s$^{-1}$)\\ [0.5mm] \hline
\multicolumn{4}{l}{{\bf WASP-173:}}\\  
57278.67586  & $-$7.3294  & 0.0220  & $-$0.0142  \\
57606.72588  & $-$8.5150  & 0.0197  & $-$0.0191  \\
57613.69950  & $-$8.4979  & 0.0143  &  0.0097 	 \\
57625.65795  & $-$7.4262  & 0.0178  & $-$0.0341  \\
57633.59834  & $-$7.4199  & 0.0233  &  0.0263 	 \\
57634.68392  & $-$8.2375  & 0.0162  & $-$0.0158  \\
57637.62434  & $-$7.8120  & 0.0174  & $-$0.0004  \\
57638.68631  & $-$8.5166  & 0.0168  & $-$0.0422  \\
57639.77874  & $-$8.1162  & 0.0141  & $-$0.0480  \\
57650.61128  & $-$7.4410  & 0.0153  & $-$0.0459  \\
57654.61358  & $-$7.2387  & 0.0231  & $-$0.0011  \\
57655.71970  & $-$7.5596  & 0.0169  & $-$0.0044  \\
57659.63846  & $-$8.2932  & 0.0164  &  0.0310 	 \\
57689.62544  & $-$7.9254  & 0.0139  & $-$0.0421  \\
57691.54377  & $-$8.2317  & 0.0172  & $-$0.0330  \\
57716.56890  & $-$8.0514  & 0.0242  &  0.0915 	 \\
57718.59164  & $-$7.5169  & 0.0225  & $-$0.0307  \\
57735.55690  & $-$8.3275  & 0.0173  & $-$0.0024  \\ [0.5mm] 
\hline
\multicolumn{4}{l}{Bisector errors are twice RV errors} 
\end{tabular}


\bsp	
\label{lastpage}
\end{document}